\documentclass[aps,prc,showpacs]{revtex4-1}

\usepackage{graphics}
\usepackage{amsmath}
\def\nuc#1#2{\relax\ifmmode{}^{#1}{\protect\text{#2}}\else${}^{#1}$#2\fi}
\newcommand{\etal}{\textit{et al.~}}

\begin{document}

%\preprint{}

\title{Exploring continuum structures with a pseudo-state basis}


%\author{}

\author{J.A. Lay}
\email{lay@us.es}
%\affiliation{Departamento de FAMN, Facultad de F\'{\i}sica, Universidad de Sevilla,
%Apdo.~1065, E-41080 Sevilla, Spain\\
%}%
\author{A.M.\ Moro}
\email{moro@us.es}
\author{J.M.\ Arias}
\email{ariasc@us.es}
\affiliation{Departamento de FAMN, Facultad de F\'{\i}sica, Universidad de Sevilla,
Apdo.~1065, E-41080 Sevilla, Spain}
\author{J.\ G\'omez-Camacho}%
\email{gomez@us.es}
\affiliation{Departamento de FAMN, Facultad de F\'{\i}sica, Universidad de Sevilla, Apdo.~1065, E-41080 Sevilla, Spain\\}
\affiliation{Centro Nacional de Aceleradores, Avda.\ Thomas A. Edison, E-41092, Sevilla, Spain}

\vspace{1cm}

\date{\today}

%\thanks{}
%\affiliation{}


\begin{abstract}
The ability of a recently developed square-integrable discrete basis 
to represent the properties of the continuum of a two-body system is investigated. The basis is obtained performing a simple analytic local scale transformation to the harmonic oscillator basis. Scattering phase-shifts and the electric transition 
probabilities $B(E1)$ and $B(E2)$ have been evaluated for several potentials using the proposed  basis. Both quantities are found to be 
in excellent agreement with the  exact values calculated from the {\it true} scattering states.  The basis has been applied to describe the
projectile continuum in the  $^6$He scattering by  $^{12}$C and  \nuc{208}{Pb} targets at 240 MeV/nucleon 
and the \nuc{11}{Be} scattering by \nuc{12}{C} at 67 MeV/nucleon. The calculated breakup differential 
cross sections are found to be in very good agreement with the available experimental data for these reactions.
\end{abstract}

%\keywords{}

\maketitle

%-------------------------------------------------------------------
\section{\label{intro} Introduction}
%------------------------------------------------------------------
Quantum collisions involving weakly bound systems are known to be influenced by the coupling to the unbound states. In  
nuclear collisions, this was first evidenced in the pioneering study  by Johnson and Soper
\cite{Ron70}, who recognized the relevance of the breakup channels to understand deuteron induced reactions. In their survey, 
the deuteron continuum was represented by a single $s$-state. Later developments by Rawitscher \cite{Raw74} and
Austern  \cite{Aus87} helped to introduce a more realistic representation of the continuum, 
leading to the development of the Continuum-Discretized Coupled-Channels (CDCC) method. This method reduces the many-body 
problem to an effective three-body problem and expands the full three-body wave function in a selected set of continuum 
wave functions of a given pair subsystem Hamiltonian.  Projection of the Schr\"odinger equation onto the selected internal 
states gives rise to a set of coupled differential equations. The practical implementation of the method requires a discretization 
procedure, i.e., an approximation of the two-body continuum spectrum by a finite and discrete representation. 
Although not strictly necessary, it results numerically  advantageous to use for this representation a set of  $\mathcal{L}^2$ (i.e.,
square-integrable) functions. The standard method 
of continuum discretization consists on dividing the continuum into a set of energy or momentum intervals. For each interval, 
or \textit{bin}, a representative wave function is constructed by superposition of the scattering states within the interval 
(the average method). 

An alternative to the discretization method based on  bins is the  pseudo-state (PS)  method, in which  the wave functions describing the internal motion of the projectile are obtained as the eigenstates of the projectile Hamiltonian in a truncated
basis of square-integrable functions.  A variety of PS basis have been proposed in the literature for two-body continuum discretization \cite{Mat03,Per02,Mor09} and, more recently, also for the three-body continuum \cite{Mat04a,Mat04b,Mat06,manoli08}. 

In a recent work \cite{Mor09}, 
we proposed a PS method based on a Local Scale Transformation (LST) of the Harmonic Oscillator (HO) basis. The LST, adopted from a previous 
work of Karataglidis \etal \cite{Amos}, is such that it transforms the Gaussian asymptotic behavior into an exponential form,  thus ensuring the correct asymptotic behavior for the bound wave functions. The accuracy of this THO basis was tested for several 
reactions induced by deuteron and halo nuclei, showing an excellent agreement with the standard binning method, and an improved convergence rate. 

Due to their vanishing asymptotic behavior, it is not obvious that genuine continuum properties, such as the scattering phase-shifts, can be 
well described using square-integrable states. However, in this case one can make use of integral formulas, which require only the wave function 
within a finite region. Several prescriptions have been proposed in the literature to extract the phase-shifts from 
continuum-discretized states \cite{Haz70,Chad01,Suz09}. In this work, we will make use of the stabilization method of Hazi and Taylor 
\cite{Haz70,Haz71,Tay76} to show that the THO basis reproduces very well the exact phase-shifts. As an additional test of the quality of the THO basis, we will calculate several transition probabilities and 
their associated sum rules. Finally, we will apply the THO basis to calculate  the breakup  of the 
reactions  \nuc{6}{He}+\nuc{12}{C}  and \nuc{6}{He}+\nuc{208}{Pb} at 240 MeV/nucleon and \nuc{11}{Be}+\nuc{12}{C} at 67 MeV/nucleon, making use of the CDCC formalism. The calculations will  
be compared with existing experimental data for these reactions.

The work is structured as follows. In Section \ref{sec:thoamos} we review the THO method based on the parametric LST. 
In Section \ref{sec:phaseshifts} we introduce an integral formula suitable for the calculation of scattering phase-shifts with PS functions. 
In Section \ref{sec:bel} we recall some useful formulae to evaluate the dipole and quadrupole transition probabilities from the scattering 
states and from the pseudo-states. In Section \ref{sec:structure} these formulae are applied to study the continuum of the deuteron and \nuc{6}{He} nuclei. In Section \ref{sec:reactions} the method is 
applied to the scattering of \nuc{6}{He} by \nuc{12}{C} and \nuc{208}{Pb} at 240 MeV per nucleon and \nuc{11}{Be}+\nuc{12}{C} at 67 MeV/nucleon. Finally, in Section \ref{sec:summary} we summarize the main results of this work.

%---------------------------------------------------
\section{ The analytic LST \label{sec:thoamos} } 
%-------------------------------------------------
In this section, we briefly review the features of the PS basis used in this work. This basis was
originally developed in \cite{Amos} to describe the single-particle orbitals within a mean-field approach. In a later 
work \cite{Mor09}, we adopted this method to discretize the continuum of a two-body system within the context of the Continuum-Discretized 
Coupled-Channels method. 

The starting point is the HO basis in angular momentum representation. The radial part of the $n$-th HO function for a given 
partial wave $\ell$ is here denoted $\phi ^{HO}_{n, \ell}(s)$. 
%which, in terms of the Laguerre functions, are written as:
%\begin{equation}
%\label{eq:ho}
%\phi ^{HO}_{n, \ell}(s)=  L_{n}^{\ell + 1/2}(s^2) \phi_\mathrm{0,\ell}(r).
%\end{equation}
%where $\ell$ denotes de orbital angular momentum and $\phi_\mathrm{0,\ell}(r)$ is the radial part of the lowest HO eigenstate of a given $\ell$. 
These functions are orthogonal and constitute a complete set and,  therefore, they can be used to expand the eigenstates (bound and unbound) 
of an arbitrary potential. For a finite well, the bound state wave functions decay exponentially at large distances and hence the HO basis does not provide a suitable representation due to its Gaussian asymptotic form. A possible approach to overcome this limitation, while retaining the appealing properties of 
the HO basis, is to perform a local scale transformation (LST) that converts the Gaussian behavior into an 
exponential one \cite{SP88,PS91}. This gives rise to the so called Transformed Harmonic Oscillator (THO) basis. We will denote the radial part of these 
basis states as:
%--------------------------------------------------------
\begin{equation}
\label{eq:tho}
\phi ^{THO}_{n, \ell}(r)= \sqrt{\frac{ds}{dr}} \phi^{HO} _{n, \ell}[s(r)].
\end{equation}
%--------------------------------------------------------

%--------------------------------------------------------
%\begin{equation}
%\label{eq:tho}
%\phi ^{THO}_{n, \ell}(r)=\left[ s(r)\right]^{\ell-\ell_0} 
%L_{n}^{\ell + 1/2}[s(r)^2]\phi _\mathrm{0, \ell_0}(r).
%\end{equation}
%--------------------------------------------------------

Note that, by construction, the family of functions  
\( \phi ^{THO}_{n, \ell}(r) \) are orthogonal
and constitute a complete set with the following normalization:

\begin{equation}
 \int_{0}^{\infty} dr |\phi ^{THO}_{n, \ell}(r)|^{2}=1  \, .
\end{equation}

Moreover, they decay exponentially
at large distances, thus ensuring the correct asymptotic behavior
for the bound wave functions. In practical calculations a finite set
of functions (\ref{eq:tho})  
is retained, and the internal Hamiltonian of the projectile is
diagonalized in this truncated basis with $N$ states,  
giving rise to a set of  eigenvalues and their associated
eigenfunctions, denoted respectively $\left\{\varepsilon_n \right\}$ and 
$\{\varphi^{(N)}_{n, \ell}(r)\}$ ($n=1,\ldots,N$). As the basis size is increased, 
those eigenstates with negative energy
will tend to the exact bound states of the system, while eigenstates
with positive eigenvalues can be regarded as a finite representation
of the unbound states.

With the criterion given above, the LST is indeed not unique. In Ref.~\cite{Per01} the LST was defined in such 
a way that the first HO state is exactly transformed into the exact ground state wave function. Therefore, by 
construction, this wave function is exactly recovered for any arbitrary size of the basis. In a more recent 
work \cite{Mor09} we adopted the parametric form of Karataglidis \etal \cite{Amos}
%---------------------------------------------------------------------
\begin{equation}
\label{lstamos}
s(r)  = \frac{1}{\sqrt{2} b} \left[  \frac{1}{   \left(  \frac{1}{r}  \right)^m  +  \left(
\frac{1}{\gamma\sqrt{r}} \right)^m } \right]^{\frac{1}{m}}\ ,
\end{equation}
%----------------------------------------------------------------------
that depends on the parameters $m$, $\gamma$ and the oscillator length $b$. Following \cite{Mor09}, 
the oscillator length $b$ is treated as a variational parameter  used to minimize the ground state energy. 
Asymptotically, the function  $s(r)$ behaves as 
$s(r)\sim \frac{\gamma}{b} \sqrt{\frac{r}{2}}$
and hence the functions obtained by applying this LST to the HO basis behave at 
large distances as $\exp(-\gamma^2 r / 2 b^2)$. Therefore, the ratio $\gamma/b$ can be related to an effective linear momentum,
$k_\mathrm{eff}=\gamma^2 /2 b^2$, which  
governs the asymptotic behavior of the THO functions; as the ratio $\gamma/b$ increases, the radial extension of the basis decreases and, 
consequently, the eigenvalues obtained upon diagonalization of the Hamiltonian in the THO basis tend to concentrate at higher energies. Therefore, $\gamma/b$ determines the density of PS as a function of the excitation energy. This property was used in \cite{Mor09} to determine a suitable value for 
the ratio $\gamma/b$ in scattering calculations. For a more quantitative
measurement of the density of states we define the magnitude:
\begin{equation}
\label{densq}
\rho^{(N)}(k)=\sum_{n=1}^{N} \langle \varphi_{\ell}(k) | \varphi^{(N)}_{n,\ell} \rangle \, ,
\end{equation}
with $|\varphi_{\ell}(k)\rangle$ denoting the scattering wave function for a momentum $k$. 

With this definition the integral of the density with respect to the momentum is just the number of basis states, i.e.
\begin{equation}
\label{sum_densq}
\int_{0}^{\infty} dk \, \rho^{(N)}(k)=N ,
\end{equation}
regardless of the choice of the parameters of the LST.

In all the calculations presented in this work, the power $m$ is just taken as $m=4$. This was one of the choices done in Ref.~\cite{Amos} and, 
in fact, the authors of that work found a very weak dependence of the results on this parameter.

%----------------------------------------------------------------------------
\section{\label{sec:phaseshifts} Extracting the phase-shifts from the THO basis}
%----------------------------------------------------------------------------
The properties of the continuum states are completely determined by the phase-shifts. In a two-body  problem, 
the phase-shifts are readily obtained from the asymptotics of the radial part of the wave function. Ignoring spins for simplicity, 
the radial part corresponding to a partial wave $\ell$ can be written at large distances as:
%---------------------------------------------------------------------------------
\begin{equation}
\label{wfasym}
\varphi_\ell(k,r) \to \sqrt{\frac{2}{\pi}}[\cos \delta_\ell(k) F_\ell(kr) + \sin \delta_\ell(k)  G_\ell(kr)]\, ,
\end{equation}
%----------------------------------------------------------------------------------
where $F_\ell$ and $G_\ell$ are the regular and irregular Coulomb functions. If the potential is real,  the functions 
$\varphi_\ell$ as well as the phase-shifts $\delta_\ell$ are also real. 

Equation (\ref{wfasym})  
%relies on the asymptotic form of the scattering wave functions, it 
can not be applied to PS functions to extract the phase-shifts, because these functions vanish asymptotically. However, the phase-shifts can be also obtained 
from integral expressions, which require only the interior part of the wave functions. Here, we make use of the 
integral formula proposed by Hazi and Taylor \cite{Haz70,Tay76}, who applied this formula to extract the phase-shifts  in a one-dimensional scattering problem using a harmonic oscillator representation. We have generalized this formula to three-dimensional cases. The formula so obtained reads 
%----------------------------------------------------------------------------------
\begin{equation}
\label{hazi}
\tan \delta_\ell(k)= 
-\frac{\int_{0}^{\infty}  \varphi_\ell(k,r) [E-H] f(r) F_\ell(kr) dr}{\int_{0}^{\infty} \varphi_\ell(k,r) [E-H] f(r) G_\ell(kr) dr} \, .
\end{equation}
%------------------------------------------------------------------------------------
This formula can be derived following the same arguments outlined in Ref.~\cite{Haz70} for the one-dimensional case. We note 
that, if the exact wave functions are used for $\varphi_\ell(k,r)$, this expression becomes an alternative 
to Eq.~(\ref{wfasym})  to calculate the exact phase-shifts.  The function $f(r)$ appearing in Eq.~(\ref{hazi}) 
verifies the following properties:
%----------------------------------------------
\begin{equation}
f(r)  \xrightarrow{r \to \infty } 1 ;  \quad 
f(0)=f'(0)=0 \, . \quad 
%{\frac{df}{dr}\Bigr\rvert}_{r=0}=0
\end{equation}
%----------------------------------------------
Following \cite{Haz70}, we adopt the explicit form $f(r)=1-\exp(-\beta r^2)$, with $\beta>0$.  The aim of this function $f(r)$ is to avoid  
evaluating the  function $G_\ell(kr)$ at the origin, where it becomes singular. Therefore the parameter
 $\beta$ should be small enough to make $f(r)\approx0$ for distances of the order of the nuclear range. In the cases 
studied here, we  have chosen $\beta=0.01$~fm$^{-2}$.

%----------------------------------------------------------------------------
\section{\label{sec:bel} Electric transition probabilities in the PS basis} 
%----------------------------------------------------------------------------
The accuracy of the PS basis to represent the continuum can be studied by comparing the ground-state to continuum transition probability due to a given operator. Here we consider the important case of the electric dissociation of the initial nucleus $a$ into the fragments 
$b+c$.  This involves a matrix element between a bound state (typically the ground state) and the continuum states. 
%In the PS  basis, however, the continuum is represented by a finite set of square- integrable functions so we first show how to approximate the continuous $B(E1)$ with the discrete $B(E1)$ values obtained from the PS basis. 

The electric transition probability between two bound states $|(\ell_i s) j_i \rangle$ and $|(\ell_f s) j_f \rangle$ (assumed here to be normalized to unity) is given by the reduced matrix element
%------------------
\begin{equation}
 \label{bediscgen}
B(E\lambda; i \to f)=\frac{2j_f+1}{2j_i+1}\left | 
\langle (\ell_f s) j_f || \mathcal{M}(E\lambda) || (\ell_i s) j_i \rangle      \right |^2    ,
\end{equation}
%-----------------
where $\mathcal{M}$ is the multipole operator:
%---------------
\begin{equation}
\mathcal{M}(E\lambda \mu)= Z_\mathrm{eff}^{(\lambda)} e r^\lambda Y_{\lambda \mu}(\hat{r}) ,
\end{equation}
%--------------
with the effective charge
\begin{equation}
Z_\mathrm{eff}^{(\lambda)}=Z_b \left(\frac{m_c}{m_b+m_c}\right)^\lambda + Z_c \left(-\frac{m_b}{m_b+m_c}\right)^\lambda .
\end{equation}

In the case of a transition to a continuum of states,  
$|k (\ell_f s) j_f \rangle$, the preceding definition is  replaced by (see e.g.~\cite{Typ05}):
%The electric transition probability from an initial (bound) state $|(\ell_i s) j_i \rangle$ to a final (bound) continuum state  $|k (\ell_f s) j_f \rangle$ due to the electric operator is given by (see e.g.~\cite{Typ05})
%ground state $\phi_0(\mathrm{r}$ to the continuum states $\phi_k(\mathrm{r}$ is given by the expression:
%---------------------------------------------------------
\begin{eqnarray}
\label{becontgen}
\frac{dB(E\lambda)}{d \varepsilon} &  = & \frac{2 j_f +1 }{2 j_i+1} \frac{\mu_{bc} k }{(2 \pi)^3 \hbar^2}    \nonumber \\
                            & \times &  \left | \langle k (\ell_f s) j_f || \mathcal{M}(E\lambda) || (\ell_i s) j_i \rangle \right |^2 \, ,
\end{eqnarray}
%---------------------------------------------------------
with $k=\sqrt{2 \mu_{bc}\varepsilon}/\hbar$. Note that the extra factor appearing in Eq.~(\ref{becontgen}) with respect to  Eq.~(\ref{bediscgen}) is consistent with the convention $\langle k (\ell s) j   |k' (\ell s) j \rangle = \delta(k-k')$ and the asymptotic behavior of 
Eq.~(\ref{wfasym}). 

In the calculations presented in this work, we will ignore for simplicity the internal spins of the clusters and 
 hence $s=0$, $j_{i}=\ell_{i}$, $j_{f}=\ell_{f}$. In addition, we will consider only transitions from the ground state, so 
$\langle \mathbf{r} | (\ell_i s) j_i \rangle =\varphi_\mathrm{g.s.}(\mathbf{r})$ (where the index $\ell_i$ is omitted for shortness). This reduces expression (\ref{becontgen}) to: 
%---------------------------------------------------------
\begin{equation}
\label{becont}
\frac{dB(E\lambda)}{d \varepsilon} = \frac{2 \ell_f +1 }{2 \ell_i+1} \frac{\mu_{bc} k }{(2 \pi)^3 \hbar^2}
  \left | \langle \varphi_{\ell_f}(k) || \mathcal{M}(E\lambda) || \varphi_\mathrm{g.s.}  \rangle \right |^2 \, .
\end{equation}
%---------------------------------------------------------
The reduced matrix element is given by:
%-------------------
\begin{equation}
 \langle \varphi_{\ell_f}(k) || \mathcal{M}(E\lambda) || \varphi_\mathrm{g.s.}  \rangle 
= \frac{4\pi}{k} Z_\mathrm{eff} e D_{\ell_i,\ell_f}^{(\lambda)}
R_{\ell_i,\ell_f}^\lambda(k) \, ,
\end{equation}
%-----------------
%-------------------
%\begin{equation}
%\langle k (\ell_f s) j_f || \mathcal{M}(E\lambda) || (\ell_i s) j_i \rangle = \frac{4\pi}{k} Z_\mathrm{eff} e D_{\ell_i,\ell_f}^{(\lambda)}
%I_{\ell_i,\ell_f}^\lambda(k)
%\end{equation}
%-----------------
with $D_{\ell_i,\ell_f}^{(\lambda)}$ a  geometric factor \cite{Typ05}
% \begin{equation}
% D_{j_i,j_f}=
% \end{equation}
 and $R_{\ell_i,\ell_f}^\lambda (k)$ the radial integral
%-----------------
\begin{equation}
R_{\ell_i,\ell_f}^{\lambda}(k)= \int_0^{\infty} dr \, \varphi_{\ell_f}(k,r) r^\lambda \varphi_\mathrm{g.s.}(r)  .
\end{equation}
%where  $\varphi_{\ell_i}(r)$ and  $\varphi_{\ell_f}(k,r)$  are the radial parts of the initial and final wave functions, respectively.

Using a finite basis, one may calculate only discrete values for the transition probability.  According 
to Eq.~(\ref{bediscgen}), the $B(E\lambda)$ between the ground state and the  $n$-th PS is given by:
%%------------------
\begin{equation}
\label{beps}
B^{(N)}(E\lambda; g.s. \to n)=\frac{2 \ell_f+1}{2 \ell_i+1}
\left | \langle  \varphi^{(N)}_{n,\ell_f} || \mathcal{M}(E\lambda) ||  \varphi_{g.s.} \rangle      \right |^2  .
%\langle (\ell_f s) j_f || \mathcal{M}(E\lambda) || (\ell_i s) j_i \rangle      \right |^2
\end{equation}
%-----------------
In order to relate this 
discrete representation with the continuous distribution (\ref{becont}) one may use the simple approximation:
%---------------------------------------------------------
\begin{equation}
\label{bedisc}
\frac{dB(E\lambda)}{d\varepsilon} \Bigr\rvert_{\varepsilon=\varepsilon_n} \simeq \frac{1}{\Delta_n}B^{(N)}(E\lambda; \mathrm{g.s.} \to n)   ,
\end{equation}
%---------------------------------------------------------
where $\Delta_n=(\varepsilon_{n+1}-\varepsilon_{n-1})/2$ is an estimate for the energy width of the $n$-th PS.  This expression provides the 
$B(E\lambda)$ values only for the PS eigenvalues $\varepsilon_n$.

Alternatively, one may derive a continuous approximation to (\ref{becont}) by introducing the identity in the truncated PS basis, i.e.:
%------------------------------------
\begin{equation}
\label{closure}
I^{N}_{\ell} = \sum_{n=1}^{N} | \varphi^{(N)}_{n,\ell} \rangle  \langle  \varphi^{(N)}_{n,\ell} |  .
\end{equation}
%-----------------------------------
For $N \to \infty$ this expression tends to the \textit{exact} identity operator for the Hilbert space spanned by the eigenfunctions of the considered 
Hamiltonian. By inserting (\ref{closure}) into the exact expression (\ref{becont}) we obtain the approximated continuous distribution:
%---------------------------------------------------------
\begin{eqnarray}
\label{befold}
\frac{dB(E\lambda)}{d\varepsilon} & \simeq  & \frac{2 \ell_f+1}{2 \ell_i+1}  \frac{\mu_{bc} k }{(2 \pi)^3 \hbar^2} \nonumber \\ 
& \times &  \left | \sum_{n=1}^{N}\langle \varphi_{\ell}(k) | \varphi^{(N)}_{n,\ell} \rangle  
%B^{(N)}(E\lambda; \mathrm{g.s.} \to n) 
\langle  \varphi^{(N)}_{n,\ell} || \mathcal{M}(E\lambda) ||  \varphi_{g.s.} \rangle   \right |^2 . \nonumber \\
\end{eqnarray}
%---------------------------------------------------------
%with $B(E\lambda; 0 \to i)=|M(E\lambda; 0 \to i)  |^2$ and $\varphi_{k,\ell}$ the true scattering states. 

%It is also useful to derive sum rules for the cumulative values of the $B(E\lambda)$ distributions. The following sum rules will be tested in this work:} 
In order to test the accuracy of the THO basis to describe the continuum, we will calculate also the following 
magnitudes:

\begin{itemize}

\item The non-energy weighted sum rule:
\begin{eqnarray}
\label{newsr}
\mathrm{NEWSR} & \equiv & \int d \varepsilon \frac{dB(E\lambda)}{d\varepsilon}  \nonumber \\ 
               &  =     & \frac{2 \ell_f+1}{2 \ell_i +1} (D_{\ell_i,\ell_f}^{(\lambda)})^2 
               \langle r^{2 \lambda}  \rangle _\mathrm{g.s.}  
\end{eqnarray}
with $\langle  r^{2 \lambda}  \rangle_\mathrm{g.s.} \equiv \langle \varphi_\mathrm{g.s.} | r^{2 \lambda} | \varphi_\mathrm{g.s.} \rangle .$

\item The energy-weighted sum rule:
\begin{eqnarray}
\label{ewsr}
\mathrm{EWSR}  & \equiv & \int d \varepsilon \frac{dB(E\lambda)}{d\varepsilon} (\varepsilon - \varepsilon_\mathrm{g.s.}) \nonumber \\
               & =  & \frac{\hbar^2}{2\mu_{bc}} \lambda(2\lambda +1)\frac{2 \ell_f+1}{2 \ell_i +1}(D_{\ell_i,\ell_f}^{(\lambda)})^2   
               \langle  r^{2 \lambda-2}  \rangle_\mathrm{g.s.} \nonumber .\\
\end{eqnarray}

\item The energy-inverse weighted integral (or polarizability):
\begin{equation}
\label{polariz}
\alpha\equiv \frac{8\pi}{9}\int d \varepsilon \frac{1}{ (\varepsilon-\varepsilon_\mathrm{g.s.})}\frac{dB(E\lambda)}{d\varepsilon} .
\end{equation}

\end{itemize}

Due to their respective weight factors, the EWSR and the polarizability are useful quantities to test the accuracy of the basis to describe high-energy and 
low-energy part of the spectrum, respectively. Note that the closed expression for the EWSR is only valid for angular momentum independent Hamiltonians.  We note also that  there is no closed expression for the polarizability, but in order to calculate this 
quantity with the desired accuracy,  as well as the EWSR for angular momentum dependent Hamiltonians, one can 
directly evaluate (\ref{polariz}) using the exact continuum states integrated up to a sufficiently high excitation energy.  

%  $\ell$-independent potential and will test our basis with the focus on the high energy contributing part of the continuum. Therefore, in order to know as well how accurate is our basis in the low energy part we also calculate the energy-inverse weighted integral or polarizability:

%-----------------------------------------------------
%\section{\label{sec:calculos} Numerical applications}
\section{\label{sec:structure} Application to nuclear structure}
%-----------------------------------------------------
\subsection{Application to the deuteron}
As an illustration of the expressions derived in the preceding sections, we first consider the case of the $p$-$n$ system with a central 
potential. Following \cite{Mor02},
the interaction between the proton and the neutron is parametrized in terms of the Poeschl--Teller potential,
%------------------------------------------------
\begin{equation}
V_{pn}(r)=-\frac{V_0}{\cosh(a r)^2} \, ,
\end{equation}
%------------------------------------------------
with $V_0=102.706$~MeV and $a=0.9407$~fm$^{-1}$. With these values, the ground state energy is 2.2245 MeV, in agreement with the experimental value.

\begin{figure}
{\par\centering \resizebox*{0.45\textwidth}{!}{\includegraphics{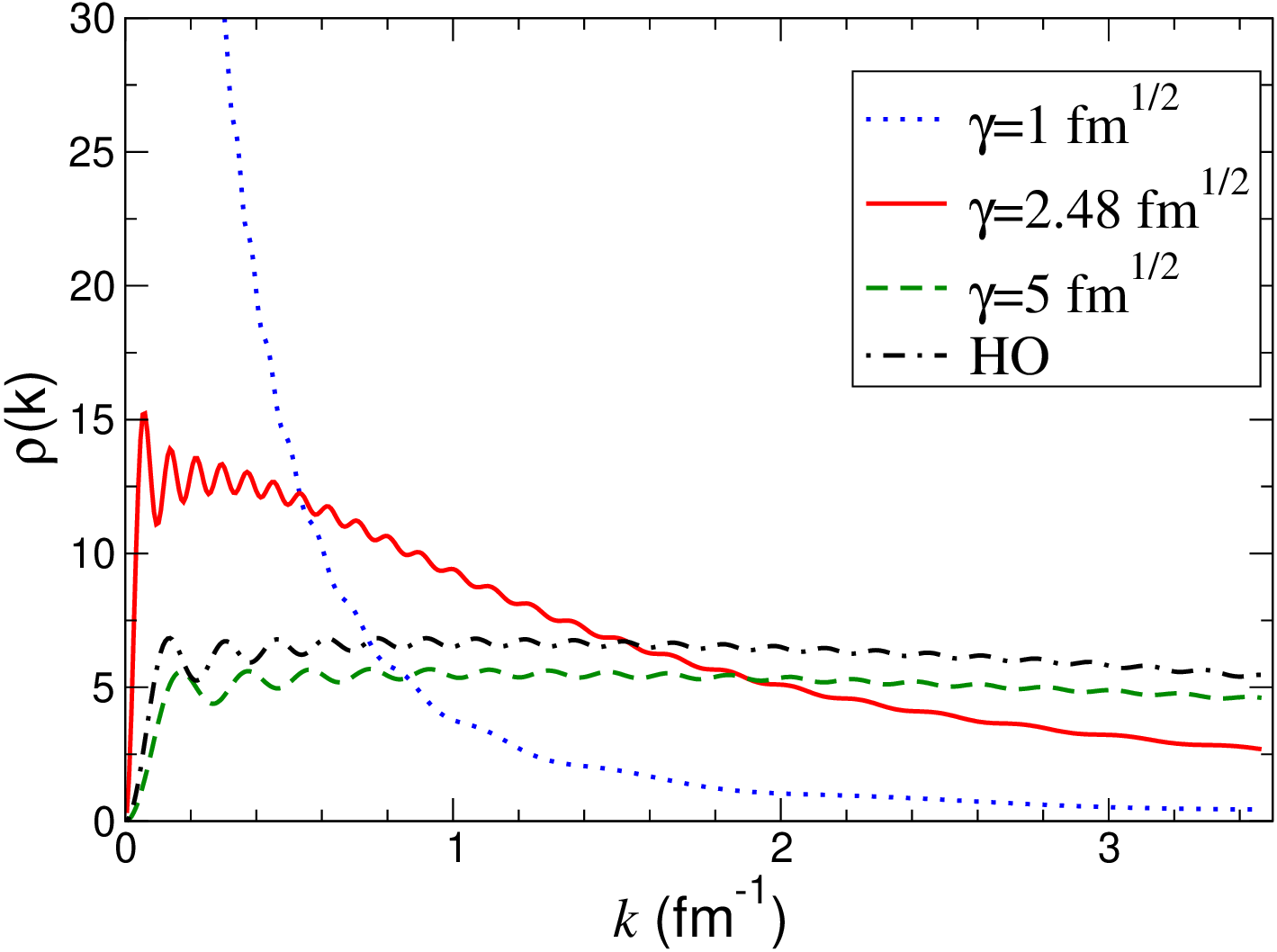}}\par}
 \caption{\label{fig:densq_deut} (Color online) Density of states for the $\ell=0$ continuum obtained with a THO basis with $N=30$ states, defined according to Eq.~(\ref{densq}) for different choices 
of the $\gamma$ parameter. The dot-dashed curve is the  density obtained with a HO basis with 30 states and $b=2$~fm.
 See text for details.}
\end{figure}

The oscillator length was chosen in order to minimize the ground state energy obtained upon diagonalization of the Hamiltonian in a small 
THO basis. This yields the value $b=1.5$~fm.  Once the value of $b$ is fixed, the ratio $\gamma/b$ determines the extension of the PS eigenstates; increasing the value of $\gamma$ reduces the radial extension and pushes the 
eigenvalues to higher energies. This is better seen in terms of the density of states, defined according to Eq.~(\ref{densq}). This magnitude is plotted in Fig.~\ref{fig:densq_deut} for the $\ell=0$ continuum, using  a basis of $N=30$ states, and three different choices of  $\gamma$, namely, $\gamma$=1~fm$^{1/2}$, 2.48~fm$^{1/2}$ and 5~fm$^{1/2}$. It is seen that small values of $\gamma$ (which correspond to an extended THO basis in configuration 
space)  produces a fine description of the continuum at low energy. This is useful, for example, to study Coulomb breakup. Increasing the value of
 $\gamma$  will decrease the density of states at low energies, that is compensated by an increase of the density 
at higher excitation energies. The most suitable choice for this parameter will depend on the problem at hand, depending on the energy region of interest. We emphasize,  however, that the dependence on $\gamma$ is not critical and, in the applications shown here, different values of $\gamma$ converge to the same results for sufficiently large bases. For comparison, we include also in Fig.~\ref{fig:densq_deut}  the density obtained with a HO basis with $N=30$ states and  $b=2.0$ fm, which minimizes the ground-state energy for the HO 
basis.

%--------------------------------------------------------------------------------------------------------------------
%------- 

\begin{figure}
{\par\centering \resizebox*{0.5\textwidth}{!}{\includegraphics{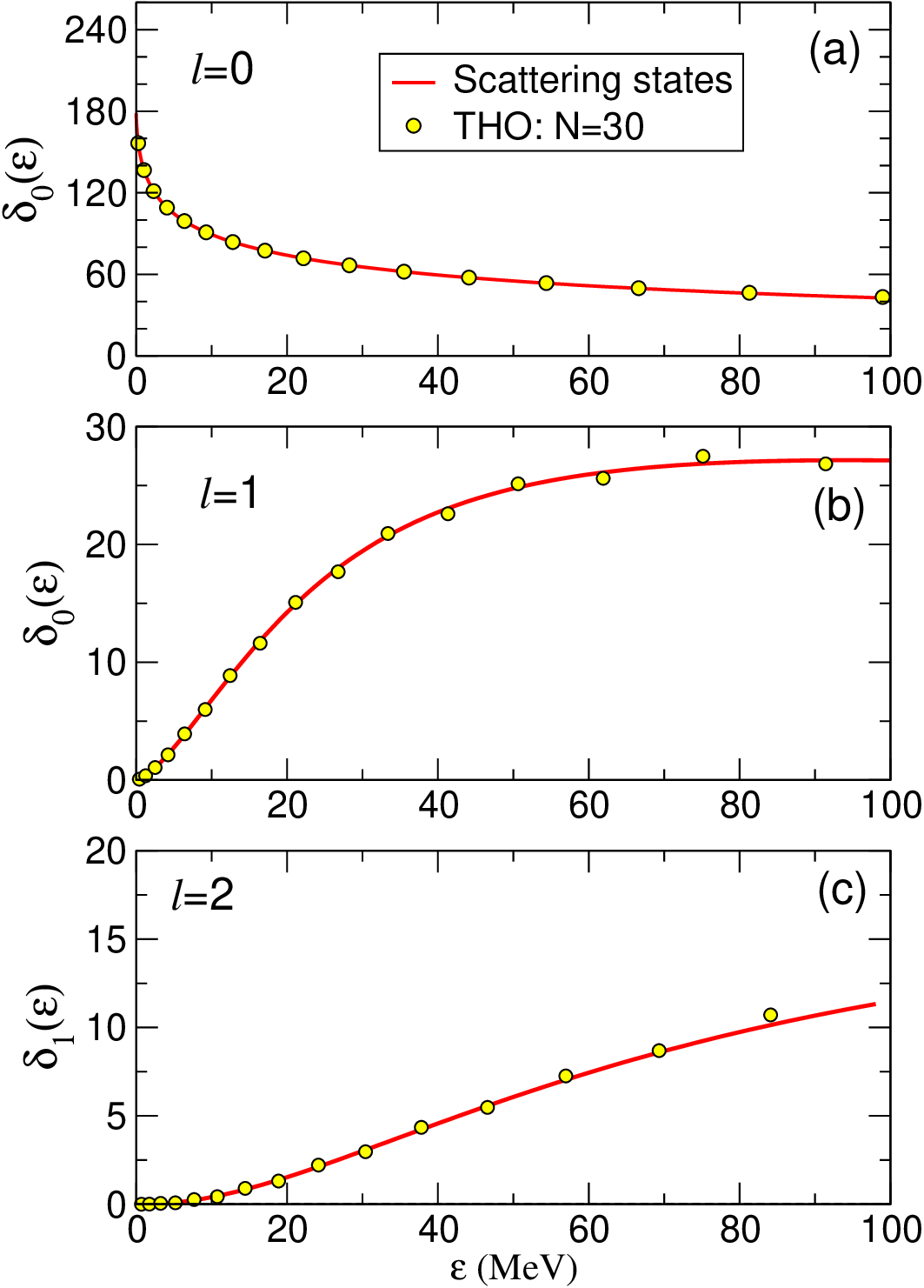}}\par}
 \caption{\label{deut-delta} (Color online) Phase-shifts for the deuteron system as a function of the relative $p$-$n$ energy. 
The upper, middle and bottom panels correspond  to $\ell$=0, 1  and 2 continuum states, respectively. See text for details.}
\end{figure}

We next consider the scattering phase-shifts. These are displayed in Fig.~\ref{deut-delta} as a function of the relative $p$-$n$ energy. The top, 
middle and bottom panels correspond to the $s$, $p$ and $d$-waves. The solid line is the calculation using the asymptotics of the exact scattering states, whereas the circles represent the calculation obtained with Eq.~(\ref{hazi}), using a THO basis with $N=30$ states. The LST was 
generated with the parameters $b=1.5$~fm and $\gamma=2.48$~fm$^{1/2}$. In the three cases, we find an excellent agreement between the \textit{exact} and approximate phase-shifts in the whole energy range. 

Note that the calculated phase-shifts are consistent with the  Levinson theorem (see e.g.~\cite{Joa87}), which establishes that the phase-shift at zero energy is given by $\delta_\ell(0)= n \pi$,  where $n$ is the number of bound states for the partial wave $\ell$. So, since the 
$s$-wave supports a bound state (the deuteron), we have $\delta_0(0)=\pi$, whereas for $\ell=1$ and $\ell=2$ we have $\delta_1(0)=\delta_2(0)=0$.  

%--------------------------------------------------------------------------------------------------------------------
%--------------------------------------------------------------------------------------------------------------------

%--------------------------------------------------------------------------------------------------------------------
\begin{figure}
{\par\centering \resizebox*{0.45\textwidth}{!}{\includegraphics{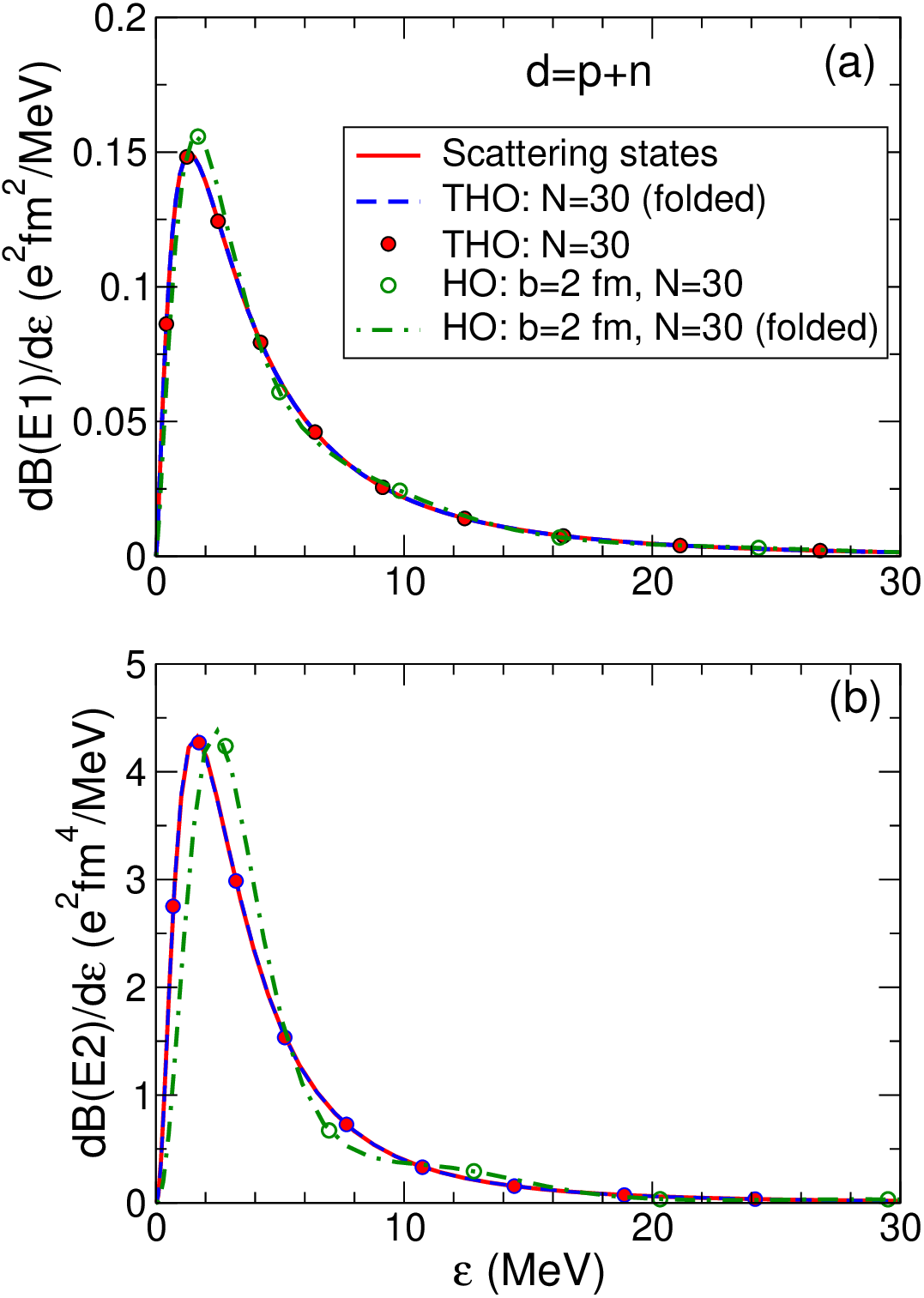}}\par}
 \caption{\label{deut-bel} (Color online) Electric transition probabilities for the  $d$=$p$+$n$ system.
The upper and bottom panels correspond  to $\lambda=1$ and $\lambda=2$ transitions, respectively. See text for details.}
\end{figure}
%------- 
We now consider the electric transition probabilities, $B(E1)$ and $B(E2)$. These are shown in Fig.~\ref{deut-bel}. The solid line 
corresponds to the calculation using the scattering states [Eq.~(\ref{becont})], the filled circles correspond to the discrete approximation 
using the THO basis [Eq.~(\ref{bedisc})], and the dashed line is the calculation obtained folding the discrete distribution with the continuum 
states [Eq.~(\ref{befold})]. Both the discrete and folded approximations show an excellent agreement with the exact distribution. We include 
also the calculation using the HO basis with $N=30$ states and $b=2.0$~fm (open circles and dot-dashed line).  Both the $B(E1)$ and $B(E2)$ distributions depart significantly from the exact distributions. In addition, the HO basis produces a small density of states at low energy, which might be a drawback for scattering calculations. 

%\begin{itemize}
%\item
%\textbf{Deberiamos incluir alguna respuesta para l=0, eg, $O(r)=v(r)$ , $O(r)=r^2$, etc ??}
%\item 
%\textbf{Mostramos la convergencia con la base de HO?}
%\end{itemize}

In Tables \ref{deut-sumrule} and \ref{deut-sumrule2} we present the convergence of the ground state energy and the $E1$ and $E2$ sum rules with respect to the basis size. The last row lists the exact values 
%for the ground state energy and for the sum rules, 
obtained with the closed expressions of Eqs.~(\ref{newsr},\ref{ewsr}). It is seen that with a 
moderately small basis one obtains a very good convergence to the exact values.  For comparison, in this Table we include also the calculations using the HO basis. From the quoted numbers, it is clear that the convergence rate is much faster for the THO basis.

It is worth noting that, despite the simple Hamiltonian adopted in this work for the $p-n$ system, 
%we do not expect that the calculated values reproduce accurately the experimental data. Nevertheless, 
the calculated polarizability is fully consistent with the experimental value $\alpha_\mathrm{exp}=0.61\pm0.04$, quoted in \cite{Fri83}.

%---------------------------------------------------------------------------
\begin{table*}%[H] add [H] placement to break table across pages
\caption{\label{deut-sumrule} Convergence of the ground state energy and the total $B(E1)$ and $B(E2)$ transition 
probabilities for the deuteron case.}
\begin{ruledtabular}
\begin{tabular}{ccccccc}
$N$ &  \multicolumn{2}{c}{$\varepsilon_\mathrm{gs}$ (MeV)} &  \multicolumn{2}{c}{Total $B(E1)$ (e$^2$fm$^2$) } & \multicolumn{2}{c}{Total $B(E2)$ (e$^2$fm$^4$)} \\
%    &    (MeV)                   &    (e$^2$fm$^2$)   &                           \\ 
    &    HO    &       THO       &     HO            &     THO         &     HO       &     THO \\
\colrule
10  &    -2.1570  &  -2.2150   &  0.81380   &    0.85286    &   14.6993     & 17.9729      \\ 
20  &    -2.2201  &  -2.2245   &  0.86200   &    0.87129    &   19.1257     & 20.7938      \\
30  &    -2.2237  &  -2.2245   &  0.86926   &    0.87136    &   20.2919     & 20.8297      \\
40  &    -2.2241  &  -2.2245   &  0.87079   &    0.87136    &   20.6422     & 20.8297      \\
50  &    -2.2242  &  -2.2245   &  0.87119   &    0.87136    &   20.7591     & 20.8297         \\
60  &    -2.2243  &  -2.2245   &  0.87130   &    0.87136    &   20.8015     & 20.8297       \\ 
70  &    -2.2243  &  -2.2245   &  0.87134   &    0.87136    &   20.8178     & 20.8297       \\ 
Exact &  \multicolumn{2}{c}{\textbf{-2.2245}}     &    \multicolumn{2}{c}{\textbf{0.87136}}  &  \multicolumn{2}{c}{\textbf{20.8297}}      \\
\end{tabular}
\end{ruledtabular}
\end{table*}

\begin{table}%[H] add [H] placement to break table across pages
\caption{\label{deut-sumrule2} Convergence of the polarizability and EWSR for the deuteron case. }
\begin{ruledtabular}
\begin{tabular}{ccccc}
$N$ &  \multicolumn{2}{c}{$\alpha$ } &   \multicolumn{2}{c}{Energy weighted $B(E1)$  }  \\
    &    \multicolumn{2}{c}{(fm$^3$)}   &    \multicolumn{2}{c}{(e$^2$fm$^2$MeV)}           \\ 
    &    HO    &       THO       &     HO            &     THO          \\
\colrule
10  &   0.413988   & 0.515397  &  8.605586    &  8.102094    \\ 
20  &   0.557520   & 0.619350  &  7.742321    &  7.442755          \\
30  &   0.599683   & 0.620899  &  7.526588    &  7.437322         \\
40  &   0.613204   & 0.620922  &  7.466005    &  7.434683       \\
50  &   0.617923   & 0.620931  &  7.446418    &  7.433209               \\
60  &   0.620109   & 0.620936  &  7.439133    &  7.432331                             \\ 
70  &   0.620395   & 0.620940  &  7.435989    &  7.431757             \\
 Exact  &  \multicolumn{2}{c}{\textbf{0.620953}}      &  \multicolumn{2}{c}{\textbf{7.429937}}    \\
\end{tabular}
\end{ruledtabular}
\end{table}
%------------------------------------------------------------------------------

%---------------------------------------------------------------------------
%\begin{table}%[H] add [H] placement to break table across pages
%\caption{\label{deut-sumrule} Convergence of the ground state energy and the total $B(E1)$ and $B(E2)$ transition probabilities for the deuteron case.}
%\begin{ruledtabular}
%\begin{tabular}{cccc}
%$N$ &   $\varepsilon_\mathrm{gs}$   &    Total $B(E1)$   & Total $B(E2)$ \\
%    &    (MeV)                   &    (e$^2$fm$^2$)   &  (e$^2$fm$^4$) \\ 
%\colrule
%10  &    -2.2150                 &    0.8028          &    14.4059      \\ 
%20  &    -2.2245                 &    0.8707          &    20.6534      \\
%30  &    -2.2245                 &    0.8714          &    20.8304      \\
%40  &    -2.2245                 &    0.8714          &    20.8316      \\
%Exact &  \textbf{-2.2245}        &    \textbf{0.8714}          &    \textbf{20.8301}      \\
%\end{tabular}
%\end{ruledtabular}
%\end{table}
%-----------

%--------------------------------------
\subsection{Application to \nuc{6}{He}}
%--------------------------------------
We now consider a situation in which more complicated continuum structures are present, such as resonances. For this purpose, 
we take the \nuc{6}{He} nucleus, treated as a two-body system $\alpha$+$2n$. Following 
\cite{Mor07}, the interaction between the two clusters is described with a Woods-Saxon shape, with $R=1.9$~fm and $a=0.39$~fm. For  
$\ell=0$ states, the depth of this potential is adjusted to give the effective separation energy of 1.6~MeV between the two clusters. 
It was shown in Ref.~\cite{Mor07} that using this effective binding energy, instead of the two-neutron separation energy ($S_{2n}=0.97$ MeV), provides a more 
realistic description of the ground state wave function.
 For  $\ell=2$, the inter-cluster potential 
is adjusted to yield a resonance at  an excitation energy of $E_x=1.8$~MeV. For $\ell=1$, we simply took the depth found for $\ell=0$. A THO basis with $N=50$ states was used, and the LST  was generated with the parameters $b=1$~fm and  $\gamma=1.89$~fm$^{1/2}$.
%--------------------------------------------------------------------------------------------------------------------
\begin{figure}
{\par\centering \resizebox*{0.5\textwidth}{!}{\includegraphics{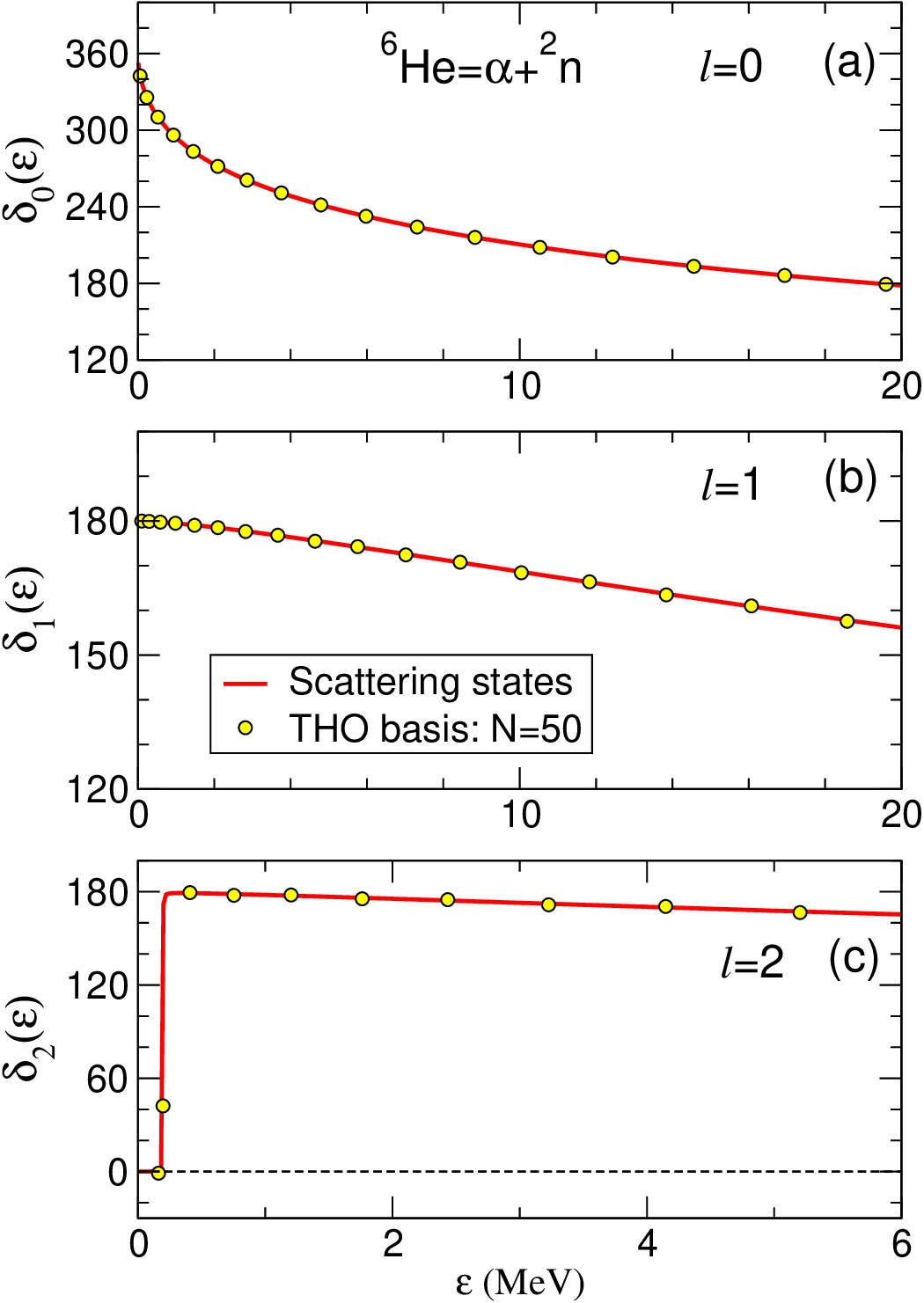}}\par}
 \caption{\label{6he-delta} (Color online) Phase-shifts for the \nuc{6}{He} system as a function of the relative $\alpha$+2n energy. 
The upper, middle and bottom panels correspond  to $\ell$=0, 1  and 2 continuum states, respectively. See text for details.}
\end{figure}
%---------------------------------------------------------------------------------------------------------------------

\begin{figure}
{\par\centering \resizebox*{0.5\textwidth}{!}{\includegraphics{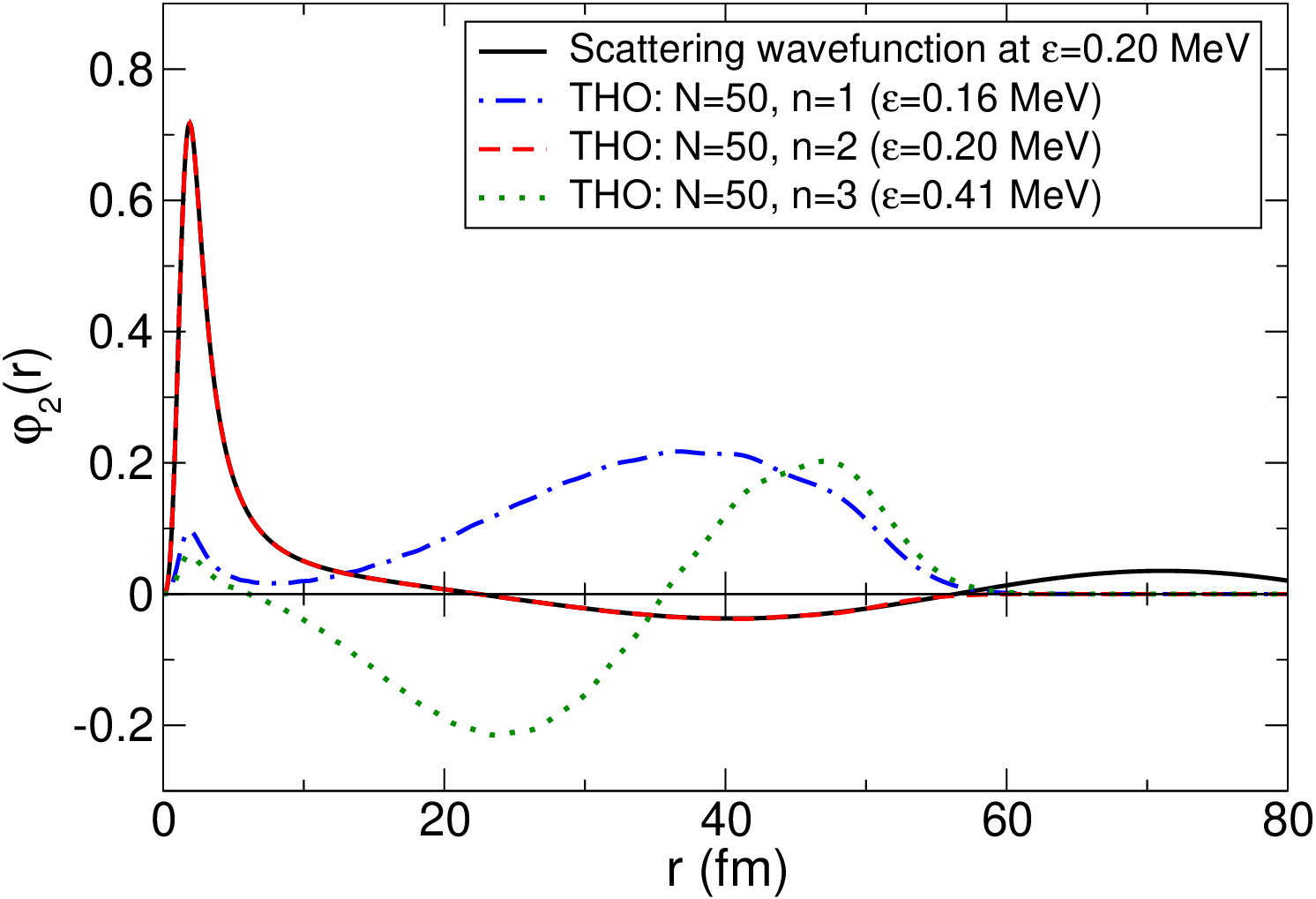}}\par}
 \caption{\label{wfres} (Color online) Radial part of the continuum wave function at the resonance for the \nuc{6}{He} system.}
\end{figure}

The $s$-, $p$- and $d$-wave phase shifts are displayed in Fig.~\ref{6he-delta} as a function of the $\alpha$+2n relative 
energy, $\varepsilon$. For $\ell=2$, the energy scale has been restricted to the energy interval $\varepsilon=0-6$~MeV in order to emphasize the 
region of the resonance.   Again, we find a perfect  
agreement between the exact (solid line) and approximate (circles) phase-shifts in the whole energy range. Note that, in this case, the 
$s$-wave potential supports two bound states, the Pauli forbidden 1S state and the 2S ground state. Consistently, the phase-shift 
at zero energy is given by $\delta_0(0)=2\pi$. Analogously, for $\ell=1$ the phase-shift tends to $\pi$, due to the presence of a (Pauli forbidden) bound state in this partial wave. Finally, for $\ell=2$ no bound states are supported by this potential and, therefore, 
$\delta_2(0)=0$. For the $d$-wave, the phase-shift crosses abruptly $\pi/2$ at $\varepsilon =0.20$~MeV, reflecting the presence of a narrow $2^+$ resonance. Interestingly, this behavior is also observed in the THO basis, where there is a PS that appears exactly at the nominal 
energy of the resonance. It is then tempting to conclude that this PS will carry most of the character of the resonance, and in fact this is confirmed in Fig.~\ref{wfres}, where we show the radial part of the exact scattering wave function, 
calculated at the energy of the resonance (solid line) along with the radial part of the PS eigenstate that appears at the energy of the resonance 
(dashed line). The former has been arbitrary normalized in order the two wave functions coincide at the maximum. It is seen that both wave functions are very similar up to 
very large distances. For comparison, we have included also the PS eigenstates associated to the eigenvalues 
just below (dotted-dashed line) and above (dotted line) the resonant one.  They are very different 
from the scattering wave function at the resonance. In particular, it can be seen that a significant part of the norm of the resonant wave function is
concentrated in the interior, as expected for a resonance, whereas for the non-resonant PS eigenstates, the probability in the interior is very small. This leads to the conclusion that 
in this case the character of the resonance is very well described by a single PS eigenstate. Indeed, if the basis is increased, the resonant character will be distributed among several PS.  These results clearly show that a distinctive feature of the 
continuum such as the resonant structures are very well accounted for by the PS basis,  despite its wrong asymptotic behaviour. 
%--------------------------------------------------------------------------------------------------------------------
%---------

% Transition probability 
%--------------------------------------------------------------------------------------------------------------------
\begin{figure}
{\par\centering \resizebox*{0.45\textwidth}{!}{\includegraphics{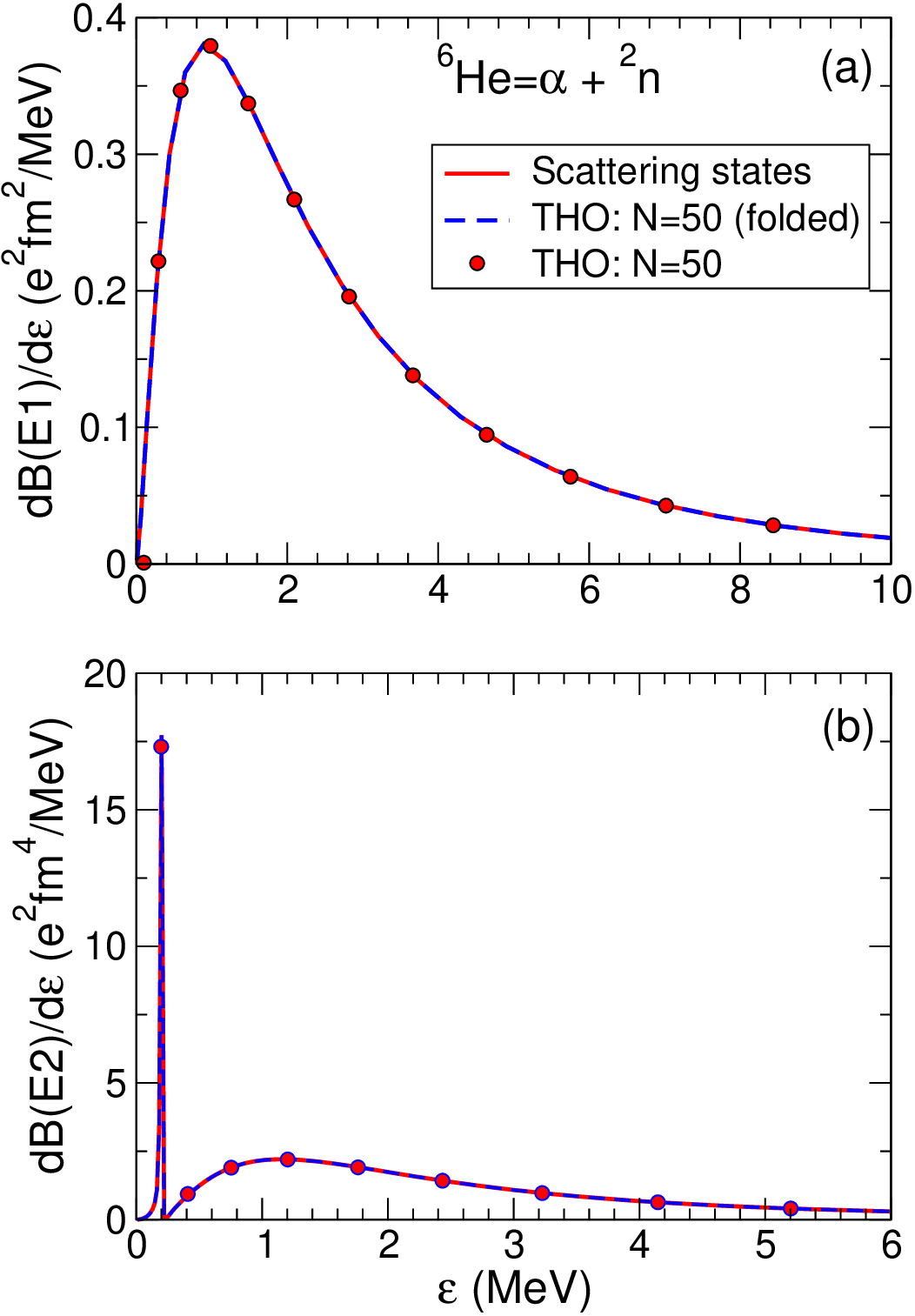}}\par}
 \caption{\label{6he-be12} (Color online) Electric transition probabilities for the   \nuc{6}{He}=$\alpha$+2n system.
The upper and bottom panels correspond  to $\lambda=1$ and $\lambda=2$ transitions, respectively. See text for details.}
\end{figure}
%----------------------------------------------------------------------------------------------------------------------

In Fig.~\ref{6he-be12} we compare the $E1$ and $E2$ transition probabilities   obtained from the scattering states by means of 
Eq.~(\ref{becont}) (solid line), with 
the approximate distributions calculated with the THO basis. The circles correspond to the discrete expression of Eq.~(\ref{bedisc}), 
whereas the dashed line is the smooth distribution obtained with Eq.~(\ref{befold}). Both the discrete and smooth distributions are in excellent agreement with the exact distribution.

 In Table \ref{he6-sumrule} we display the convergence of the ground state energy and the $E1$ and $E2$ sum rules with respect to the basis size. As in the deuteron case, both  observables converge very fast to their exact value, given by the sum rule of Eq.~(\ref{newsr}).

Finally, in Table \ref{he6-sumrule2} we present the convergence of the integrated energy weighted  $B(E1)$  and the polarizability. For the 
former, we can not use the closed expression of Eq.~(\ref{ewsr}), because the \nuc{6}{He} Hamiltonian depends on the angular momentum. The 
\textit{exact} value listed  in this Table corresponds to the explicit calculation using the scattering states up to a high excitation energy. The same
holds for the polarizability, since for this observable there is no closed expression.

%---------------------------------------------------------------------------
\begin{table}%[H] add [H] placement to break table across pages
\caption{\label{he6-sumrule} Convergence of the ground state energy and the total $B(E1)$ and $B(E2)$ transition 
probabilities for the \nuc{6}{He}=$\alpha$+2n system.}
\begin{ruledtabular}
\begin{tabular}{cccc}
$N$ &     $\varepsilon_\mathrm{gs}$  &  Total $B(E1)$   & Total $B(E2)$ \\
    &     (MeV)                      & (e$^2$fm$^2$)   &  (e$^2$fm$^4$) \\ 
\colrule   
% 5  &        -1.39148      &             &          2.22077  \\
10  &        -1.5913      &   1.3538    &           8.5524       \\
20  &        -1.5999      &    1.3854   &           9.7471     \\
30  &        -1.5999      &    1.3855    &          9.7471     \\
40  &        -1.5999      &    1.3855    &          9.7471     \\
50  &         -1.6000     &     1.3855    &           9.7471     \\
80  &         -1.6000     &     1.3855    &           9.7471     \\
Exact &  \textbf{-1.6000}  &    \textbf{1.3855}   &       \textbf{9.7471}         \\
\end{tabular}
\end{ruledtabular}
\end{table}
%------------------------------------------------------------------------------
%\begin{table}%[H] add [H] placement to break table across pages
%\caption{\label{he6-sumrule} Convergence of the ground state energy and the total $B(E1)$ and $B(E2)$ transition 
%probabilities for the \nuc{6}{He}=$\alpha$+\nuc{2}{n} system.}
%\begin{ruledtabular}
%\begin{tabular}{cccc}
%$N$ &     $\varepsilon_\mathrm{gs}$  &  Total $B(E1)$   & Total $B(E2)$ \\
%    &     (MeV)                      & (e$^2$fm$^2$)   &  (e$^2$fm$^4$) \\ 
%\colrule   
% 5  &        -1.39148      &             &          2.22077  \\
%10  &        -1.59324      &   1.2011    &           5.94853       \\
%20  &        -1.59995      &    1.2495   &           7.81096      \\
%30  &        -1.59997      &    1.2748    &          7.85382     \\
%40  &        -1.59999      &    1.2748    &          7.85578     \\
%50  &         -1.60000     &     1.2748    &           7.85947     \\
%80  &         -1.60000     &     1.2748    &                \\
%Exact &  \textbf{-1.6000}  &    \textbf{1.2748}   &       \textbf{7.854227}         \\
%\end{tabular}
%\end{ruledtabular}
%\end{table}
\begin{table}%[H] add [H] placement to break table across pages
\caption{\label{he6-sumrule2} Convergence of the polarizability and energy weighted $B(E1)$ for the \nuc{6}{He}=$\alpha$+2n system.
% The experimental value for the polarizability is taken from \cite{Pac07}.
}
\begin{ruledtabular}
\begin{tabular}{ccc}
$N$ &  $\alpha$  &  Total E. W. $B(E1)$  \\
    &    (fm$^3$)  &    (e$^2$fm$^2$MeV)        \\ 
\colrule
20  &   1.8652   &  6.5623       \\ 
30  &   1.8746   &  6.5464             \\
40  &   1.8750   &  6.5438             \\
50  &   1.8752   &  6.5425         \\
60  &   1.8753   &  6.5418                 \\
70  &   1.8753   &  6.5412                               \\ 
120 &   1.8755   &  6.5401               \\
Exact &  \textbf{1.8756}   &    \textbf{6.5393}   \\
%Experimental &  \textbf{1.99$\pm$0.40}     & Exact &  \textbf{$\ell$-dependent potential}   \\
\end{tabular}
\end{ruledtabular}
\end{table}
%------------------------------------------------------------------------------

\section{\label{sec:reactions} Applications to nuclear reactions}

%---------------------------------------------------------------------------------------------------
\subsection{Application to the reactions \nuc{6}{He}+\nuc{12}{C} and  \nuc{6}{He}+\nuc{208}{Pb}  at 240 MeV/nucleon}
%----------------------------------------------------------------------------------------------------
%One of the purposes of the THO basis proposed in this work is to provide a suitable representation of the continuum 

The THO basis considered  in this work is intended to provide a suitable discrete representation of the continuum spectrum of a loosely 
bound system, which can be useful for scattering calculations within 
%The main purpose of the PS developed in this work is its application as a discretization procedure within
 the Continuum-Discretized 
Coupled-Channels method \cite{Aus87}. As a test case, we will apply the THO basis to the reactions 
\nuc{6}{He}+\nuc{12}{C} and \nuc{6}{He}+\nuc{208}{Pb}  at 240 MeV/nucleon. 
These reactions were measured 
by Aumann \etal \cite{Aum99} at the GSI facility with the aim of  extracting information on the  \nuc{6}{He} nucleus. The breakup of  
\nuc{6}{He} on \nuc{208}{Pb} was already analysed using the CDCC method with the THO basis in Ref.~\cite{Mor09}, showing an excellent agreement with the binning discretization method  for the modulus of the breakup $S$-matrix. In this work, 
we extend the analysis of \cite{Mor09} in order to compare with the  data of Ref.~\cite{Aum99}. In particular, we  consider the exclusive breakup differential cross section as a function of the excitation energy of the projectile  $d\sigma/dE_x$. In Ref.~\cite{Aum99}, this observable was obtained by reconstructing the kinematics of the 
$^6$He c.m.\ from the measured momenta of the outgoing fragments ($^4$He+n+n) and integrating up to a laboratory scattering angle of 80 mrad. To obtain this observable in our calculations, we first construct the double differential cross section $d\sigma/d\Omega dE_x$ from the breakup $S$-matrices. In principle, the breakup $S$-matrix is a continuous function of the asymptotic momentum $k$. However, within a PS representation of the continuum, only discrete values of the $S$-matrix are obtained, corresponding to the eigenvalues $\varepsilon_n$. A continuous 
breakup $S$-matrix can be obtained from the solution of the coupled equations 
% and then integrate in the c.m. angle of the $^6$He. Th
following the procedure used in \cite{Mat03,Tos01,Mor09}, in which the discrete $S$-matrices are folded with the exact scattering states, similarly to 
what was done with the $B(E\lambda)$ distribution in Eq.~(\ref{befold}), i.e.
%----------------------------------------------
\begin{equation}
\label{scont}
S_{f:i}(k) 
\approx
\sum_{n=1}^{N} 
\langle \varphi_{\ell_f}(k) | \varphi^{(N)}_{n, \ell_f} \rangle 
\hat{S}^{(N)}_{n:i}(k_n) ,
\end{equation}
%-------------------------------------------------------- 
where $\hat{S}^{(N)}_{n:i}(k_n)$ are the discrete $S-$matrix elements resulting from 
the solution of the coupled--channels equations using a PS basis with $N$ states. The subscripts $i$, $n$ and $f$ denote the 
channels 
$\{\varphi_\mathrm{g.s.}; L_i, \ell_i, J\}$,  $\{\varphi^{(N)}_{n, \ell_f}; L_f, \ell_f, J\}$ and $\{\varphi_{\ell_f}(k) ; L_f, \ell_f, J\}$, respectively,
where $L_i$ ($L_f$) is the  
initial (final)  orbital angular momentum for the projectile--target relative motion, and $J$ the total angular momentum of the  system.
%In this method, the breakup $S-$matrix elements $S_{\alpha':\alpha}(k)$, which depend  
% on the continuous variable $k$, as well as on the initial and final
% angular momenta,  are obtained by an appropriate superposition of  the discrete
% $S-$matrix elements   $\hat{S}_{\alpha':\alpha}(k_i)$ resulting from 
% the solution of the coupled--channels equations, as \cite{Mat03,Tos01}

%----------------------------------------------------------------
\begin{figure}[thb]
{\par\centering \resizebox*{0.45\textwidth}{!}
{\includegraphics{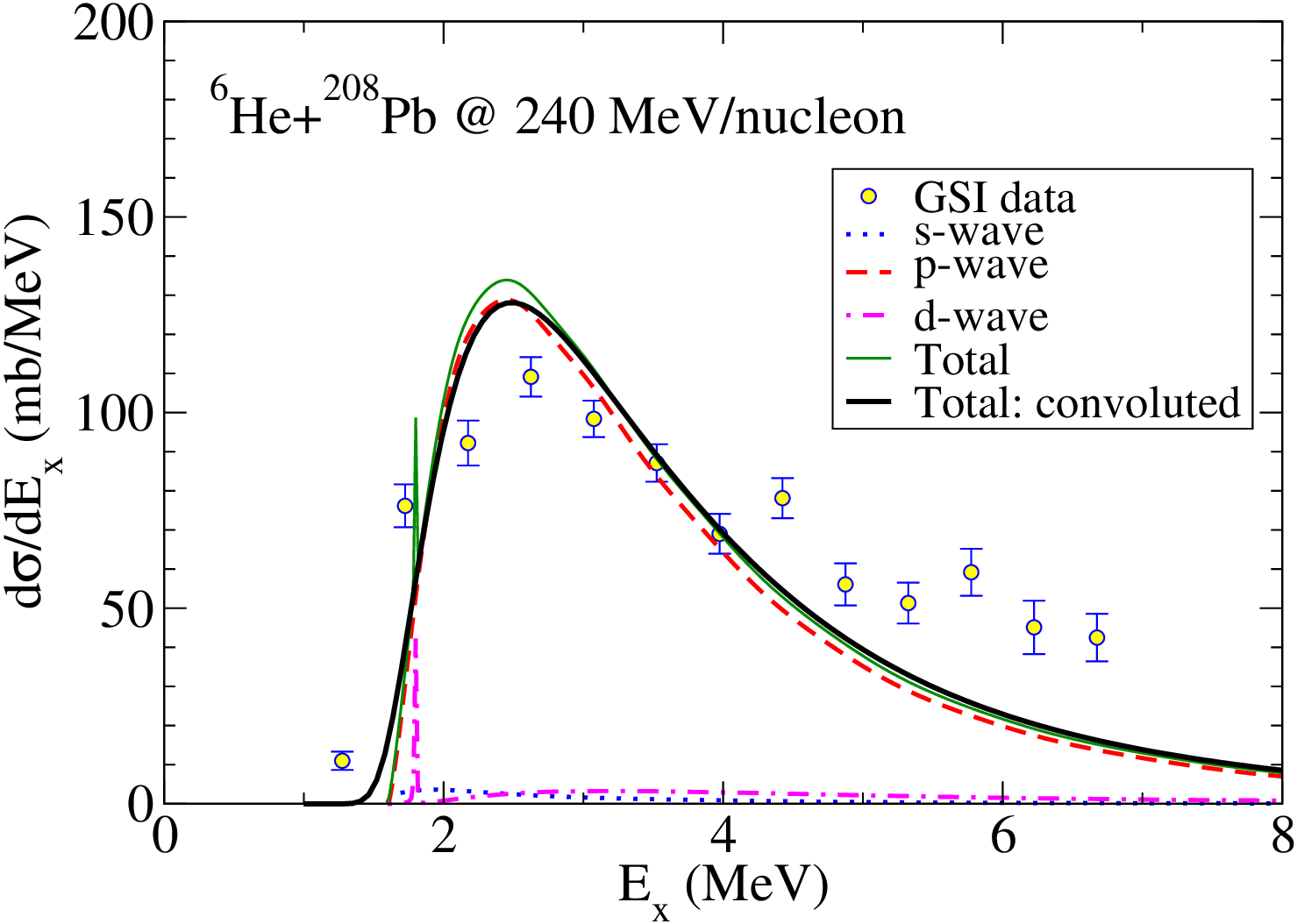}}\par}
\caption{\label{he6pb_dsde} (Color online) Angle-integrated breakup differential cross section, as
a function of the excitation energy, for the
reaction \nuc{6}{He}+\nuc{208}{Pb}  at 240 MeV/nucleon. The dotted, dashed and dotted-dashed 
lines are the contribution of the $s$, $p$ and $d$ states. The thin solid line is the sum of these contributions and the thick solid line is the full calculation convoluted with the experimental resolution.  The filled circles are the data from Ref.~\cite{Aum99}.
}
\end{figure}
%-----------------------------------------------------------------

% Pb target
We first consider the Pb target. 
For the $2n$-$\alpha$ potential and the parameters of the THO basis  we keep the same values of the preceding section.  These parameters were found to provide an appropriate distribution of eigenstates for the \nuc{6}{He}+\nuc{208}{Pb} reaction at  240 MeV/nucleon \cite{Mor09}.  The $\alpha$+\nuc{208}{Pb} and $2n$+\nuc{208}{Pb} interactions are needed to generate the diagonal and non-diagonal coupling potentials of the CDCC equations. For the $\alpha$+\nuc{208}{Pb} interaction we have adopted the first optical potential used in Ref. \cite{Bon85}.
 For the $2n$+\nuc{208}{Pb} interaction, we used the following single-folding model:
\begin{equation}
U(\mathbf{R})= \int \rho_{nn}(r_{nn}) \{U_n (\mathbf{R}+\frac{\mathbf{r}_{nn}}{2}) + U_n(\mathbf{R}-
\frac{\mathbf{r}_{nn}}{2})   \} d\mathbf{r}_{nn}
\end{equation}
where $U_n$ is the neutron-\nuc{208}{Pb} optical potential taken from the parametrization of Madland \cite{Mad97}, evaluated at the apropriate energy per nucleon, and $\rho_{nn}(r)$ is the neutron-neutron density distribution. The latter was calculated 
integrating the square of the three-body wave function of the \nuc{6}{He} nucleus along the $2n$-$\alpha$ coordinate, i.e.\
%\begin{equation}
%\label{eq_rho3b}
%\rho(r_{nn})= y^2 \int |\Psi^{3B}(\mathbf{r}_{nn},\mathbf{y})|^2 
%d\mathbf{y} d\Omega_{nn} \,\, ,
%\end{equation}
\begin{equation}
\label{eq_rho3b}
\rho(r_{nn})= r_{2n-\alpha}^2 \int |\Psi^{3b}(\mathbf{r}_{nn},\mathbf{r}_{2n-\alpha})|^2 
d\mathbf{r}_\mathrm{2n-\alpha} d\Omega_{nn} \,\, ,
\end{equation}
where $\Psi^{3b}(\mathbf{r}_{nn},\mathbf{r}_{2n-\alpha})$ 
is the three-body wave function  and 
$\Omega_{nn}$ denotes the angular variables $(\theta_{nn}, \phi_{nn})$. In the present calculations, the function $\Psi^{3b}$ was taken from Ref.~\cite{Mor07}.

Very good convergence of the CDCC calculations was achieved with a basis of $N=30$ states. In addition, we found 
that  continuum states above 50 MeV have a negligible effect on the scattering observables, and hence these eigenstates were removed from the coupled-channels calculation. This reduces the number of PS  included in the CDCC equations to $n_s=n_p=14$,  $n_d=15$ for $s$, $p$ and $d$ waves. The coupled equations were integrated up to 100~fm, and for a total angular momentum up to $J_\mathrm{max}=2000$. 

In Fig.~\ref{he6pb_dsde} we show the calculated energy differential breakup cross section, along with the GSI data. The dotted, dashed, and dotted-dashed lines are the separate contributions of the $s$, $p$ and $d$ continuum states. For this angular 
range ($\theta_\mathrm{lab} < 80$~mrad) the cross section is largely dominated by the coupling to the $j=1^-$ states due, mainly, to the strong 
dipole Coulomb interaction. 

 The thin solid line in Fig.~\ref{he6pb_dsde} is the sum of the $s$, $p$ and $d$ contributions. For a meaningful comparison with the data, this curve has to be convoluted with the experimental energy resolution, which we took from the same work \cite{Aum99}. The result of this folding is shown by the thick solid line. At low excitation energies, this calculation reproduces very well the shape and magnitude of the data. For excitation energies $E_x>4$~MeV, the calculation underpredicts the data. This discrepancy was also found in the semiclassical calculations 
reported in \cite{Aum99}. Note that the narrow peak due to the $2^+$ resonance in $^6$He disappears in the folded calculation. 
%Similar experiments have been recently developed 
%with \nuc{11}{Be} and \nuc{11}{Li} projectiles. In this work, we do not attempt to compare with these data, since this would require the introduction of more realistic interactions for the fragment-target systems, and the convolution of our results with the experimental response. 

The dominance of the dipole Coulomb couplings at these very small angles was used in Ref.~\cite{Aum99} 
to extract the $dB(E1)/d\varepsilon$ distribution, by comparing the measured differential cross section, 
$d\sigma/dE_x$, with semiclassical calculations. Although our calculations confirm the dominance of the $E1$ 
couplings, we have found that nuclear potentials have a small but not negligible effect on this observable. In 
addition,  starting from the same structure model for 
the \nuc{6}{He} nucleus, our CDCC calculations show some 
departure from the semiclassical calculations, suggesting that the connection between the energy differential cross 
section and the underlying $E1$ probability is more complicated than suggested by the semiclassical approach. These results are potentially very interesting because  they may affect the extracted $dB(E1)/d\varepsilon$  distribution from the cross section data. This analysis is beyond the scope of the present work and then we leave it for a separate publication. 

% The fact that the differential breakup cross section 
%$d\sigma/dE_x$ is mainly determined by the  electric transition probability $dB(E1)/d\varepsilon$, a fact that was used in Ref.~\cite{Aum99} to extract this transition probability by comparing the experimental cross section with 
%theoretical calculations based on a semiclassical framework. Our calculations indicate that the nuclear potentials 
%play a small role, although not negligible on this observable.  This results could be of relevance for  Since this discussion is beyond the scope of the present paper, we leave  aim of this work is just to show the usefulness of the THO basis to describe the breakup channels, we leave this discussion for a 

\begin{figure}[t]
{\par\centering \resizebox*{0.45\textwidth}{!}
{\includegraphics{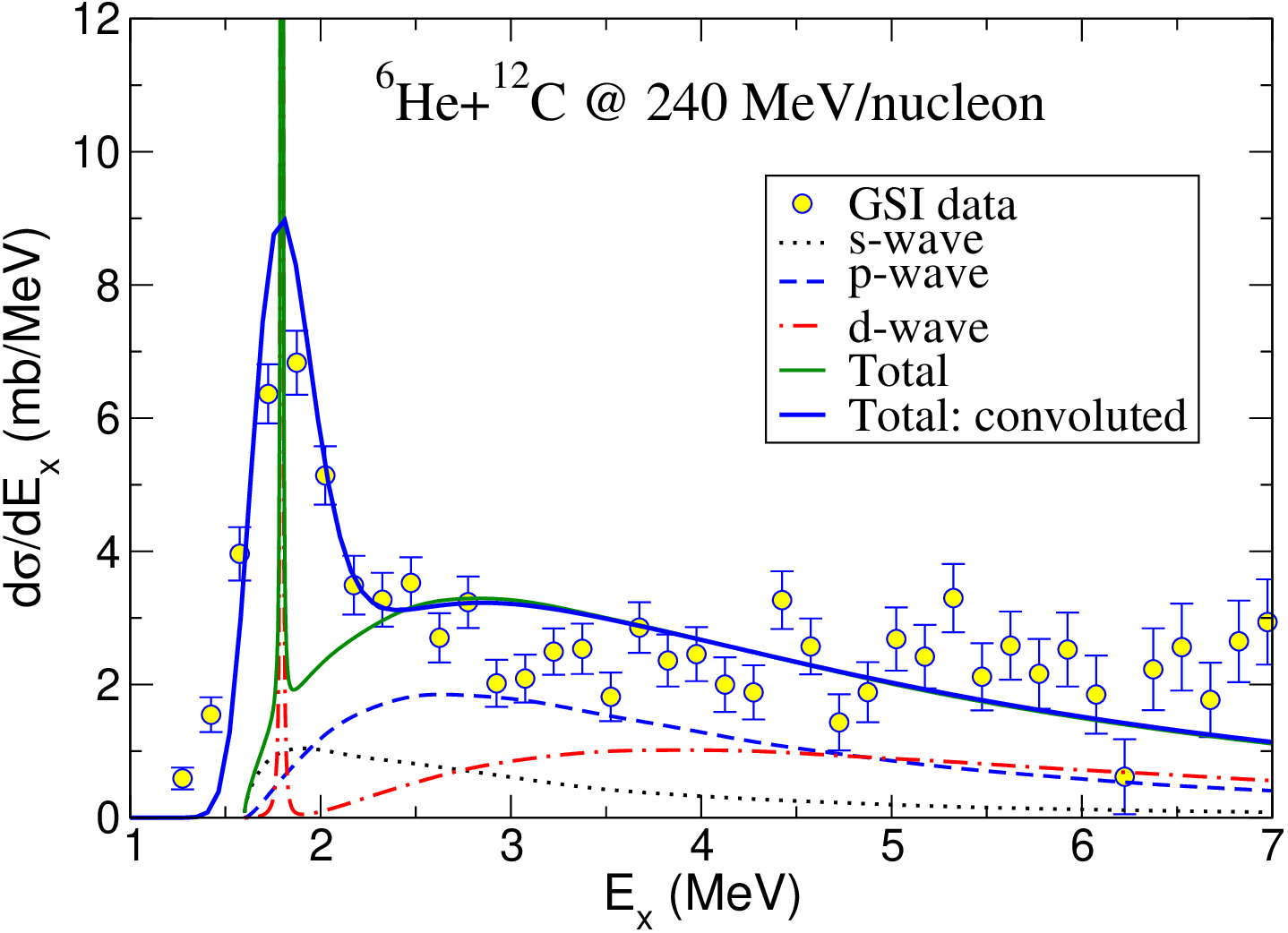}}\par}
\caption{\label{he6c_e240A_dsde} (Color online) Energy breakup differential cross section, as
a function of the excitation energy, for 
reaction \nuc{6}{He}+\nuc{208}{Pb}  at 240 MeV/nucleon. The dashed line is the CDCC calculation with the THO discretization method. The solid line is the result of convoluting the CDCC calculation with the experimental resolution. The filled circles are the data from Ref.~\cite{Aum99}. 
}
\end{figure}

% 12C target
We consider now the \nuc{12}{C} target.  The $2n$-$\alpha$ interaction and the parameters of the LST where the same used in the Pb case.  For the \nuc{6}{He} continuum states,  $\ell=0-3$ waves were included. In this case, we used a THO basis with   $N=40$ states. The $\alpha$+\nuc{12}{C} potential was adopted from 
Ref.~\cite{Kho02}. For the $2n$+\nuc{12}{C} interaction we used the same single-folding model as for the \nuc{208}{Pb} taking also the neutron-target  potential, $n$+\nuc{12}{C} in this case, from the parametrization of Madland \cite{Mad97}.
The coupled equations were integrated up to 100~fm, and for a total angular momentum up to $J_\mathrm{max}=200$. In Fig.~\ref{he6c_e240A_dsde} we compare the CDCC calculations with the experimental data from \cite{Aum99}. The meaning of the curves is the same as in Fig.~\ref{he6pb_dsde}. In this case, the low energy cross section is dominated by the population of the $2^+$ continuum, with $s$ and $p$ waves contributing to the background. The contribution of the $f$-wave was found to be very small and hence it has not been included in this figure. As in the Pb case, the $d$-wave cross section shows a narrow peak corresponding to the well known resonance at $E_x=1.8$~MeV. The width of the peak in our calculation is significantly smaller than the experimental  width of this resonance ($\Gamma_\mathrm{exp} \approx 100$~keV). This is a consequence of our simple two-body model adopted for the $^6$He nucleus. Nevertheless, when the calculation is convoluted with the experimental resolution (thick solid line) it becomes very close to the data. Despite the simplicity of the structure model adopted in these calculations, these results show that the THO discretization method constitutes a useful method to 
describe accurately detailed structures of the continuum that may show up in scattering observables.

%----------------------------------------------------------------
%-------

 As a final remark on these results, we note that, at these energies, relativistic effects might affect the calculated observables. Some recent exploratory studies \cite{Oga09,Oga10} have shown that these effects produce an increase of
about 10-15 \% on the 
calculated breakup cross section for the \nuc{8}{B}+\nuc{208}{Pb} and \nuc{11}{Be}+\nuc{208}{Pb} reactions at 250 MeV/nucleon. These effects
affect only the very small angles and are mostly due to the modification of the Coulomb potential. Therefore, we might expect a similar effect in our \nuc{6}{He}+\nuc{208}{Pb} case. For the \nuc{6}{He}+\nuc{12}{C} case, we do not expect these effects to be important, because dynamical relativistic corrections to 
the nuclear interaction were found by the same authors to be negligible at these energies.  These corrections refer to dynamical effects exclusively. Relativistic kinematics effects were included in the referred works, as well as in our calculations, by using the appropriate relativistic momentum.  In any case, the aim of our work is to show the ability of the THO basis to describe the continuum of a two-body system, and so the emphasis of our study is more on the description of the structure, 
rather than on the reaction mechanism.

%---------------------------------------------------------------------------------------------------
\subsection{Application to the \nuc{11}{Be}+\nuc{12}{C} reaction at 67 MeV/nucleon}
%----------------------------------------------------------------------------------------------------
As a final example, we consider the scattering of the halo nucleus  \nuc{11}{Be} on a carbon target. This 
reaction has been recently measured by Fukuda and collaborators at RIKEN \cite{Fuk04}, in order to extract information 
on the \nuc{11}{Be} continuum by measuring neutron-\nuc{10}{Be} coincidences following the projectile breakup. The angle-integrated differential cross section as a function of the relative energy between the outgoing neutron and \nuc{10}{Be}, displays a structure dominated by a prominent resonance 
at $E_x = 1.78$~MeV. This resonance was interpreted as a $d_{5/2}$ neutron coupled to the \nuc{10}{Be} ground state. A second bump was also observed at $E_x=3.41$~MeV,  which was given a tentative assignment $3/2^+$, with a 
small contribution of the $d_{3/2}$ wave coupled to \nuc{10}{Be}($0^+$) and a larger contribution of the
\nuc{10}{Be}($2^+$)$\otimes \nu 2s_{1/2}$ configuration. As in previous analyses of this reaction \cite{Cap04,How05}, 
we use a two-body model of the projectile, \nuc{11}{Be}=\nuc{10}{Be}(g.s.)+n, and hence those states based on the 
core excited states are absent from our model-space.  

In the present CDCC calculations, the neutron-\nuc{10}{Be} interaction was taken from \cite{Cra10}. This potential reproduces the  separation energy of the ground state ($1/2^+$) and first excited state ($1/2^-$), and the position of the $5/2_1^+$ resonance, assuming the configurations $2s_{1/2}$, $1p_{1/2}$ and 
$d_{5/2}$, respectively. The \nuc{11}{Be} continuum was described with a THO basis with $N=25$ states. The LST  was generated with the parameters $b=2.4$ fm and $\gamma=3.6$ fm$^{1/2}$.  Continuum states with configuration $s_{1/2}$, $p_{1/2}$, $p_{3/2}$, $d_{3/2}$ and $d_{5/2}$  were considered. After diagonalization of the projectile Hamiltonian in this THO basis, only those eigenstates with excitation energies below $E< 20$ MeV were included in the coupled-channels calculations, since the breakup cross section was found to be  very small above this energy. This leaves about 
10--11 eigenstates for each partial wave. Following \cite{How05},  the $n$+\nuc{12}{C} and \nuc{10}{Be}+\nuc{12}{C} potentials 
were taken from \cite{Bec69} and \cite{Alk97}, respectively. 
 The coupled-channels equations were integrated up to a matching radius of $R=90$~fm and for total angular momenta up to $J=350$.

\begin{figure}[t]
{\par\centering \resizebox*{0.45\textwidth}{!}
{\includegraphics{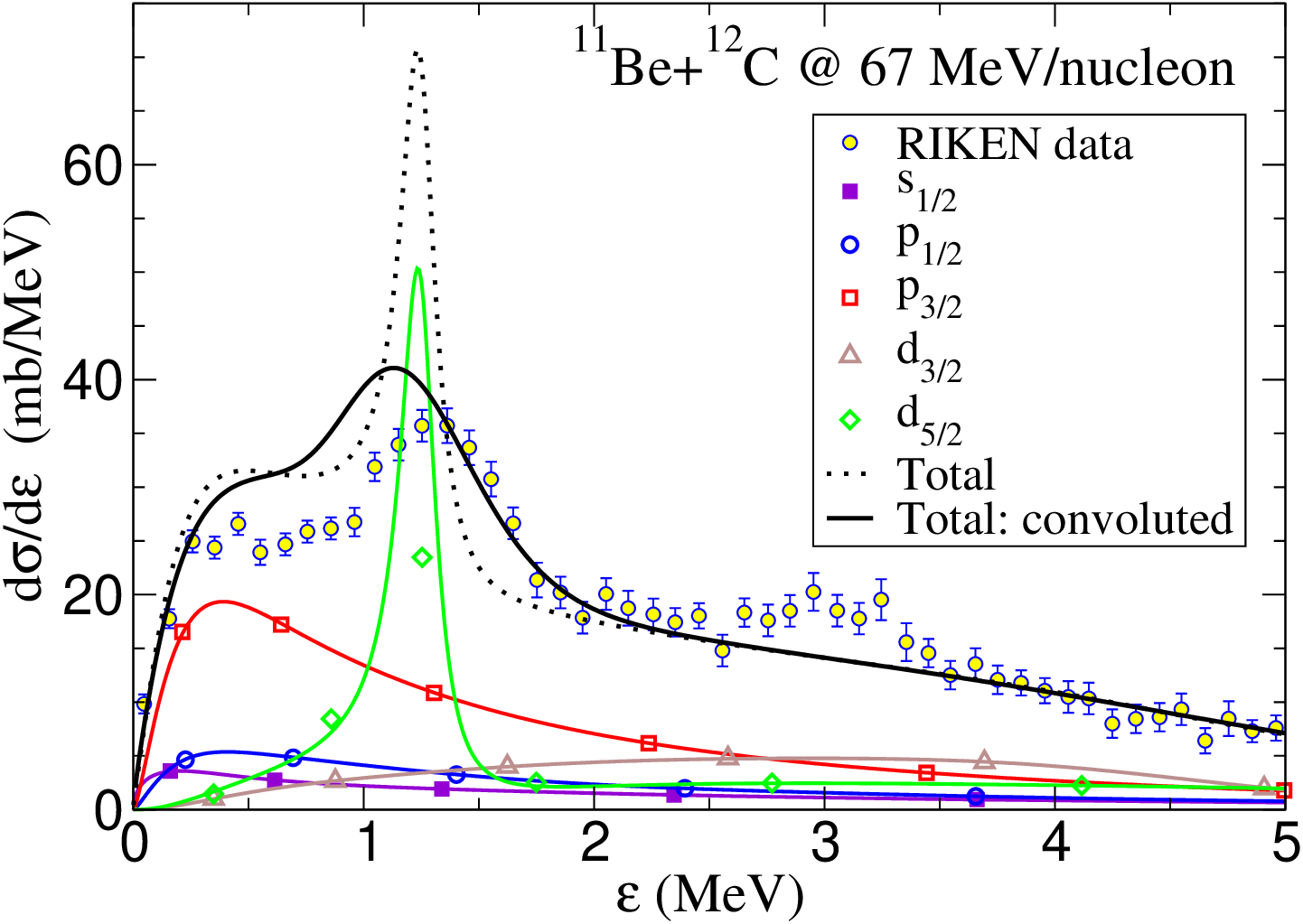}}\par}
\caption{\label{be11c_dsde} (Color online) Energy breakup differential cross section, as
a function of the excitation energy, for 
reaction \nuc{11}{Be}+\nuc{12}{C}  at 67 MeV/nucleon. The separate contribution of each partial wave is shown. The dotted line is the full contribution, and the thick solid line is the folding of the latter with the instrumental 
resolution. The experimental data are from Ref.~\cite{Fuk04}. 
}
\end{figure}

In Fig.~\ref{be11c_dsde} we compare the experimental \cite{Fuk04} and calculated energy differential cross section 
as a function of the $n$-\nuc{10}{Be} relative energy. Both the data and the calculations correspond to the 
angular range $0^\circ \leq \theta_\mathrm{c.m.}  \leq 12^\circ$. The calculated contribution of each partial wave is shown.  The symbols correspond to the contribution of specific pseudo-states, whereas the continuous lines are obtained convoluting the discrete $S$-matrices with the {\em exact} continuum states. The low-lying continuum is dominated by the $p_{3/2}$ and $d_{5/2}$ waves, with the latter being responsible for the resonant peak at $E_x=1.78$ MeV. The dotted line is the sum of the different partial waves, and the thick solid line is the result of folding this full calculation with the instrumental resolution quoted in \cite{Fuk04}. At energies close to the breakup threshold, the calculation overestimates the data. At energies above the resonance peak, the shape and magnitude of the data is well reproduced, with the exception of the broad peak due to the second resonance. We remind, however, that this resonance is believed to contain a significant contribution of \nuc{10}{Be}(2$^+$) and, therefore, it is not expected to be well described with our model space. Our results 
are very close to those obtained by Howell {\it et al.} \cite{How05}, including the overestimation of the data at low excitation energies. Since the calculations of that work used a continuum discretization in terms of energy bins, we attribute this discrepancy with the data  to the choice of the interactions or to the restrictions of our three-body model, rather than to the method of  discretization. As discussed in the \nuc{6}{He}+\nuc{12}{C} case, an advantage of the THO discretization over the standard binning method is the ability of describing fine structures of the continuum with a relatively small basis. For example, to describe $d_{5/2}$ resonance, the CDCC calculations of Ref.~\cite{How05}
used 15 bins for $\varepsilon = 0.5-2$ MeV, whereas in the present calculations about 10 PS are enough to 
describe the full energy region, including the narrow resonance.
%once the discrete energy distribution is convoluted with the continuum states. 

%----------------------------------------------------------------
%-----------------------------------------------------------------

\begin{figure}[t]
{\par\centering \resizebox*{0.45\textwidth}{!}
{\includegraphics{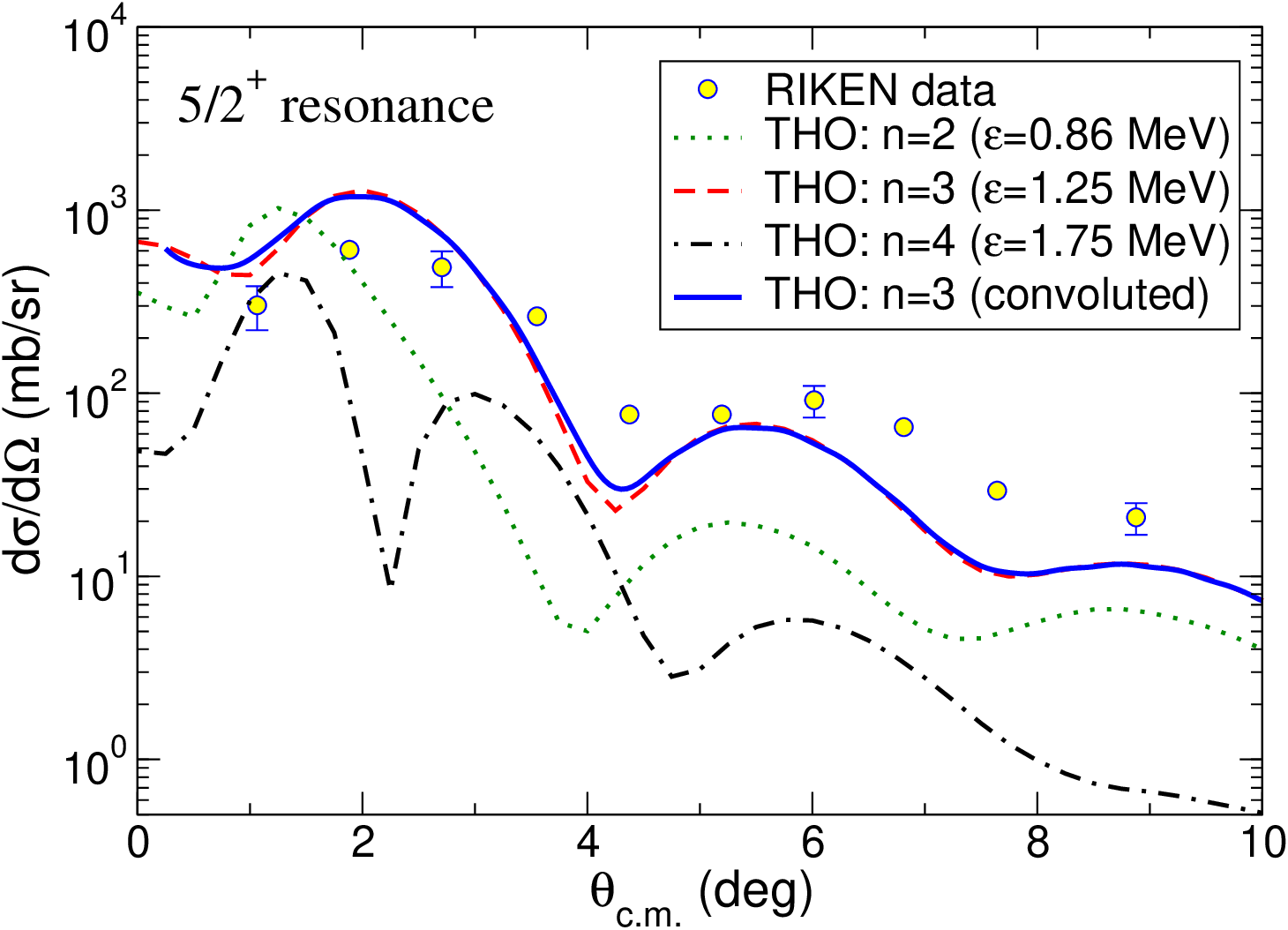}}\par}
\caption{\label{be11c_dsdw_res} (Color online) Angular distribution  for the $E_x=1.78$~MeV resonance.  The circles are the data points from Ref.~\cite{Fuk04}. The curves 
are the angular distribution due to the three PS around the resonance. See text for details. 
}
\end{figure}

As in the \nuc{6}{He} case, for the $d_{5/2}$ continuum we get an eigenstate at $\varepsilon=1.25$ MeV, which is close to the nominal energy of the resonance ($\varepsilon=1.27$ MeV) and hence it is plausible to associate this eigenstate with the resonance structure. To corroborate 
this conclusion, in Fig.~\ref{be11c_dsdw_res} we compare  the experimental angular distribution of the resonance region \cite{Fuk04} with the angular distribution of the three eigenstates closer to the nominal resonance energy. As anticipated, the 
eigenstate at $\varepsilon=1.25$ MeV reproduces fairly well the shape and magnitude of the data, supporting our conclusion that this eigenstates carries most of the resonant character. It has to be borne in mind that, as the basis size is increased, the resonant character will be distributed among several eigenstates and hence this identification is not possible. In fact, for the present basis size ($N=25$) there might be some mixing between the three eigenstates shown in Fig.~\ref{be11c_dsdw_res}.
% and this might explain the fact that the  angular distribution due to the $\varepsilon=1.27$ MeV state underestimates the data.

%----------------------------------------------------------------
%-----------------------------------------------------------------

%--------------------------------------------------------------------
\section{\label{sec:summary} Summary and conclusions}
%--------------------------------------------------------------------
In this work we have dealt with the problem of the study of the continuum properties of a weakly bound system in terms of 
basis of square-integrable functions, or pseudo-states (PS). The general idea of the PS method is to diagonalize the Hamiltonian of the two-body 
system in a truncated PS basis. The eigenstates with negative energies represent the bound states of the system, whereas those at positive energies are 
regarded as a finite and discrete representation of the continuum spectrum. Among the many possible choices of the PS basis, in this work 
we have made use of the Transformed Harmonic Oscillator (THO) basis  proposed in \cite{Amos,Mor09}, in which the PS functions are generated by applying 
a parametric Local Scale Transformation (LST) to the HO basis. The analytic form of the LST makes very simple the calculation of the PS basis. In 
addition, the radial extension of the basis and the energy distribution of the eigenvalues can be 
 controlled through the  parameters defining the LST. This permits to adapt the properties of the basis to the problem at hand. 

In order to test the accuracy of the THO basis to represent the continuum, we have evaluated the scattering phase-shifts for the deuteron  and $^6$He systems, treated as two-body systems ($p+n$ and $\alpha$+$2n$) interacting with a simple central interaction. Since the THO states vanish at large distances, the phase-shifts have been evaluated using an integral formula, 
following the prescription of  Hazi and Taylor \cite{Haz70}. In both cases, we find an excellent agreement with the {\it exact} phase-shifts, obtained from the 
asymptotic part of the scattering states. Even the sharp resonance in the $^6$He case is very well described with a small THO basis. 
As an additional test, we have evaluated the electric transition probabilities $B(E1)$ and $B(E2)$ for the same systems, finding again an excellent agreement with the results obtained with the scattering states. For this observable, a simple smoothing procedure has been proposed to provide 
a continuous  distribution ($dB(E\lambda)/d\varepsilon$) in terms of the discrete values obtained with the PS basis.

Finally, we have presented CDCC calculations for the reactions $^6$He+$^{12}$C and $^6$He+$^{208}$Pb at 240 MeV/nucleon, and $^{11}$Be+$^{12}$C at 67 MeV/nucleon for which experimental 
data exist \cite{Aum99,Fuk04}.  For the \nuc{11}{Be}+\nuc{12}{C} reaction, we have used a two-body model \nuc{10}{Be}(g.s.)+$n$ of the projectile.  In order to compare with the recent data of Fukuda {\it et.} \cite{Fuk04}, we have  calculated the breakup differential cross section as a function of the neutron-\nuc{10}{Be} relative energy. Using a relatively small THO basis, we have been able to reproduce fairly well the data, including the narrow $d_{5/2}$ resonance at $\varepsilon =1.27$~MeV. Interestingly, one of the THO eigenstates appears at an energy very close to this energy and its associated differential angular cross section reproduces fairly well the experimental angular distribution obtained for the resonance region.

For the \nuc{6}{He} reactions, we have used a simple two-body model  ($\alpha$+$2n$). %The calculated energy differential cross section, integrated for lab.\  angles $\theta_{lab.}<90$~mrad,  are in very good agreement with the experimental data of Aumann \etal \cite{Aum99}. 
Our calculations, which are parameter free, reproduce quantitatively and
qualitatively the experimental energy differential cross sections reported in \cite{Aum99}
%, integrated for  $\theta_{lab}<80$~mrad, by Aumann \textit{et al.}, 
both for the heavy target ($^{208}$Pb)  and for the light target ($^{12}$C).  Furthermore, these
calculations indicate that, for the $^6$He+$^{208}$Pb reaction at excitation energies below
$\sim$4~MeV, break-up cross sections are Coulomb dominated, with monopole and
quadrupole components contributing only about 6\%. However, for $^6$He+$^{12}$C, the
dominant component is the quadrupole, so that for excitation energies below $\sim$2.5~MeV, 
dipole and monopole components contribute about 24\%.

The role of nuclear forces and higher order effects has been investigated by comparing our full
coupled channels calculation with Coulomb calculations using the
equivalent photon model. Differences  as large as 28\% have been found, indicating the
need of performing continuum-discretized coupled-channels calculations
to extract structure information from break-up reaction data, at least in this energy regime.

The agreement between theory and experiment is very encouraging, given the simplicity of the dineutron model 
used in the present calculations. One has to bear in mind, however, that some of the details of the breakup 
distributions might be hidden due to  the energy resolution of the experiment. 
%This indicates that, despite the simplicity of the dineutron model for $^6$He,
%break-up cross sections can be reproduced, although this is associated
%to the energy resolution of the experiment. 
New measurements
with better energy resolution will be useful to test more stringently the break-up
distributions at energies closer to the threshold. An accurate description of these reactions 
will  require a realistic three-body model to describe the \nuc{6}{He} nucleus.
%In the Pb case, this observable is largely dominated by the  dipole Coulomb couplings, whereas in the 
%$^{12}$C case, the most salient feature is the prominent peak at $E_x \simeq 1.8$~MeV due to the coupling to the well known $2^+$ resonance of $^6$He. 
 In this respect, it is worth noting that the THO method used in this work can be generalized to three-body problems. This can provide a useful yet simple method to study continuum 
structures (e.g.\ resonances) in nuclei with a three-body structure  ($^{9,14}$Be, $^{6}$He, $^{11}$Li, $^8$B, etc) as well as reactions involving these nuclei. A similar approach proposed very recently, making use of a different PS basis, has been found to provide very promising results  \cite{Bar09,Kie10}.
%Work in this direction is already in progress. 

%--------------------------------------------------------------------------
\begin{acknowledgments}
%--------------------------------------------------------------------------
 This work has been partially supported by the Spanish Ministerio de
 Ciencia e Innovaci\'on under projects  FIS2008-04189,  FPA2009-07653, FPA2009-08848 and  by the 
Spanish Consolider-Ingenio 2010 Programme CPAN (CSD2007-00042). J.A.L.\ acknowledges a 
research grant by the Ministerio de Ciencia e Innovaci\'on. We are kindly grateful to T.\ Aumann for 
supplying us the experimental data in tabular form and for his help interpreting these data.

\end{acknowledgments}

\bibliography{thoamos}

\end{document}